\documentclass[aps,pra,twocolumn,showpacs,amsmath,amssymb,superscriptaddress]{revtex4-1}

\usepackage{graphicx, amsmath, bm, braket, txfonts, color, ulem, wasysym, fontawesome, txfonts,  mathtools}
\usepackage[dvipsnames]{xcolor}

\usepackage{mathrsfs}

\usepackage{amssymb}

\usepackage[colorlinks=true, citecolor=blue, filecolor=blue, linkcolor=blue, urlcolor=blue]{hyperref}

\usepackage{lineno}

\newcommand{\fR}{ f_{\mathrm{Rabi}} }

\newcommand{\fMW}{ f_{\rm{MW}} }
\newcommand{\lso}{ l_{\rm{SO}} }
\newcommand{\ldot}{ l_{\rm{dot}} }

\newcommand{\meff}{ m_{\rm{eff}} }

\newcommand{\VL}{ V_{\rm{L}} }
\newcommand{\VLP}{ V_{\rm{LP}} }
\newcommand{\VM}{ V_{\rm{M}} }
\newcommand{\VRP}{ V_{\rm{RP}} }
\newcommand{\VR}{ V_{\rm{R}} }
\newcommand{\Vi}{ V_{i} }
\newcommand{\THa}{ T_{2}^{\rm{Hahn}} }

\newcommand{\tburst}{ t_{\rm{burst}} }

\begin{document}

\title{Compromise-Free Scaling of Qubit Speed and Coherence}
\author{Miguel J. Carballido}
\email{miguel.carballido@unibas.ch}
\affiliation{Department of Physics, University of Basel, Klingelbergstrasse 82, CH-4056 Basel, Switzerland}

\author{Simon Svab}
\affiliation{Department of Physics, University of Basel, Klingelbergstrasse 82, CH-4056 Basel, Switzerland}

\author{Rafael S. Eggli}
\affiliation{Department of Physics, University of Basel, Klingelbergstrasse 82, CH-4056 Basel, Switzerland}

\author{Taras~Patlatiuk}
\affiliation{Department of Physics, University of Basel, Klingelbergstrasse 82, CH-4056 Basel, Switzerland}

\author{Pierre~Chevalier~Kwon}
\affiliation{Department of Physics, University of Basel, Klingelbergstrasse 82, CH-4056 Basel, Switzerland}

\author{Jonas~Schuff }
\affiliation{Department of Materials, University of Oxford, Oxford OX1 3PH, United Kingdom}

\author{Rahel~M.~Kaiser}
\affiliation{Department of Physics, University of Basel, Klingelbergstrasse 82, CH-4056 Basel, Switzerland}

\author{Leon C. Camenzind}\thanks{Currently at: CEMS, RIKEN, Wako, Saitama 351-0198, Japan}
\affiliation{Department of Physics, University of Basel, Klingelbergstrasse 82, CH-4056 Basel, Switzerland}

\author{Ang~Li}\thanks{Currently at: Institute of Microstructure and Properties of Advanced materials, Beijing University of Technology, Beijing, 100124, China}
\affiliation{Department of Applied Physics, TU Eindhoven, Den Dolech 2, 5612 AZ Eindhoven, The Netherlands}

\author{Natalia Ares}
\affiliation{Department of Engineering Science, University of Oxford, Oxford OX1 3PJ, United Kingdom}

\author{Erik~P.A.M~Bakkers}
\affiliation{Department of Applied Physics, TU Eindhoven, Den Dolech 2, 5612 AZ Eindhoven, The Netherlands}

\author{Stefano~Bosco}\thanks{Currently at: QuTech and Kavli Institute of Nanoscience, Delft University of Technology, Delft, The Netherlands}
\affiliation{Department of Physics, University of Basel, Klingelbergstrasse 82, CH-4056 Basel, Switzerland}

\author{J. Carlos Egues}
\affiliation{Department of Physics, University of Basel, Klingelbergstrasse 82, CH-4056 Basel, Switzerland}
\affiliation{Instituto de F\'isica de S\~{a}o Carlos, Universidade de S\~{a}o Paulo, 13560-970 S\~{a}o Carlos, S\~{a}o Paulo, Brazil}

\author{Daniel Loss}
\affiliation{Department of Physics, University of Basel, Klingelbergstrasse 82, CH-4056 Basel, Switzerland}
\affiliation{CEMS, RIKEN, Wako, Saitama 351-0198, Japan}

\author{Dominik M. Zumb\"uhl}
\email{dominik.zumbuhl@unibas.ch}
\affiliation{Department of Physics, University of Basel, Klingelbergstrasse 82, CH-4056 Basel, Switzerland}

\date{August 15, 2025}

\begin{abstract}
\noindent
Across leading qubit platforms, a common trade-off persists: increasing coherence comes at the cost of operational speed, reflecting the notion that protecting a qubit from its noisy surroundings also limits control over it. This speed-coherence dilemma limits qubit performance across various technologies. Here, we demonstrate a hole spin qubit in a Ge/Si core/shell nanowire that triples its Rabi frequency while simultaneously quadrupling its Hahn-echo coherence time, boosting the Q-factor by over an order of magnitude. This is enabled by the direct Rashba spin–orbit interaction, emerging from heavy-hole-light-hole mixing through strong confinement in two dimensions. Tuning a gate voltage causes this interaction to peak, providing maximum drive speed and a point where the qubit is optimally protected from charge noise, allowing speed and coherence to scale together. Our proof-of-concept shows that careful dot design can overcome a long-standing limitation, offering a new approach towards building high-performance, fault-tolerant qubits.
\end{abstract}
\maketitle
\begin{figure}[htb!]
\centering
\includegraphics[width=1\linewidth]{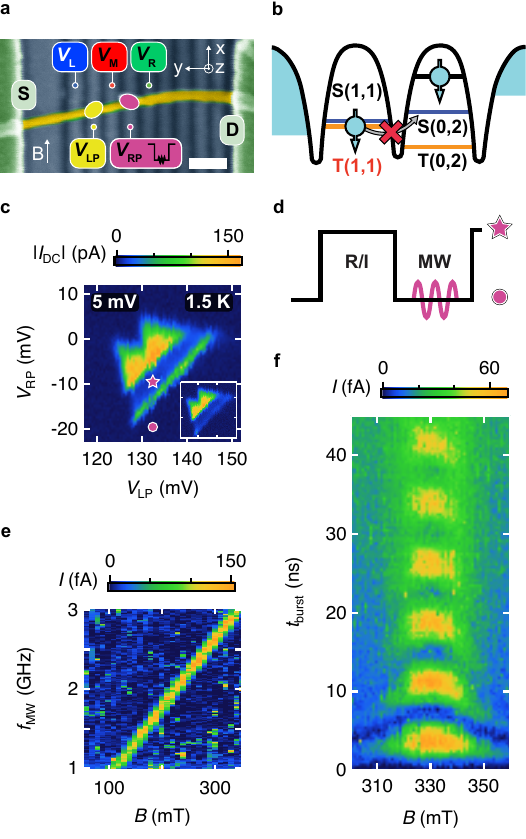} 
\caption{{\bf \textsf{Measurement of a NW hole spin qubit at 1.5~K. a,}}~False-color scanning electron micrograph of a representative NW-device. The Ge/Si NW is highlighted in yellow, lying on top of nine bottom gates, and is further connected to source and drain contacts from above, marked $S$~and~$D$. The first five bottom gates from the left were used to form a DQD, whose expected location is indicated by the yellow and pink ovals. The colour code is consistently used throughout the manuscript. The scale bar corresponds to $100$~nm. {\bf \textsf{b,}}~Schematic of PSB. In the absence of magnetic field and for detunings $\varepsilon < \varepsilon_{\rm{ST}}$, charge transport is blocked when the system is initialised in a triplet state $T(1,1)$. {\bf \textsf{c,}}~Measurement of bias triangles with lifted spin-blockade at $B=350$~mT ($x$-direction) and bias of $V_{\rm{SD}} = +5$~mV. The pink star marks the qubit initialisation/readout spot and the pink circle indicates the manipulation point. Inset: Same bias triangle at $B=~0$~mT, manifesting PSB via the suppressed current at the baseline. {\bf \textsf{d,}}~Schematic of the pulse scheme applied to $V_{\rm RP}$, consisting of the readout/initialization stage (R/I) and the manipulation stage for which a MW burst is applied while in Coulomb blockade. {\bf \textsf{e,}}~Measurement of the current as a function of $f_{\rm{MW}}$ and $\mathbf{B}$ showing characteristic EDSR at a fixed microwave burst duration. {\bf \textsf{f,}}~Current as a function of microwave burst duration and magnetic field $\mathbf{B}$ at $f_{\rm{L}} = 2.79$~GHz, showing Rabi-oscillations with a frequency $\fR = 130$~MHz and coherence time $T_2^{\rm{Rabi}} = 40$~ns.}
\label{fig:HotGeSiQubit}
\end{figure}
\begin{figure*}[htb]
\centering
\includegraphics[width=1\linewidth]{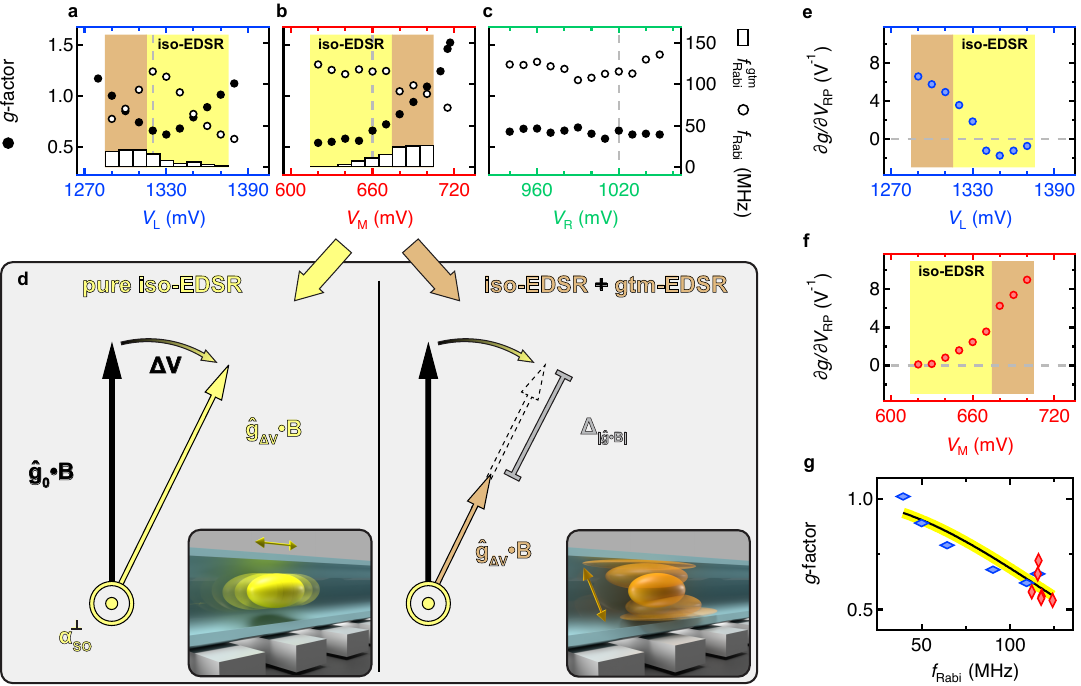} 
\caption{ {\bf \textsf{SOI mediated g-factor renormalization of an elongated dot. a-c,}}~Experimental data of the qubit g-factor (solid black discs), the corresponding Rabi frequency $\fR$ (empty discs) and the estimated gtm-EDSR contribution $\fR^{\rm{gtm}}$ (white bars), as a function of the barrier gate voltages $V_{\rm{L}}$, $V_{\rm{M}}$ and $V_{\rm{R}}$, respectively. For each of the datasets a-c, the values on the other two barrier gates are fixed at the value marked by the vertical dashed line. The yellow coloured regions highlight the voltage ranges for which the qubit driving mechanism is dominantly iso-EDSR, whereas the brown regions show the ranges for which the contribution to the measured $\fR$ originating from gtm-EDSR, $\fR^{\rm{gtm}}$, is $\geq 15\%$ ($5\%$ above the spread of data points, see Supplementary Information). These colours are used consistently in all other panels. The estimated $\fR^{\rm{gtm}}$ is represented by the white bar charts. {\bf \textsf{d,}}~Schematic visualisation of how the Zeeman vector $\mathbf{\hat{g}}_0 \cdot \mathbf{B}$ is affected when subject to an infinitesimal voltage change $dV$ in the presence of SOI represented by $\alpha^{\perp}_{\rm{SO}}$ and a fixed magnetic field $|\mathbf{B}|$. Left shows pure iso-EDSR  coming from displacements of the dot potential. Right shows the additional appearance of gtm-EDSR, when the dot is displaced and deformed. {\bf \textsf{e,}}~Shows the extracted $\partial g/\partial \VRP$, as a function of the voltage $\VL$ in blue, and {\bf \textsf{f,}}~as a function of $\VM$ in red. These derivatives are used to calculate an upper bound of the $g$-modulated contribution to $\fR$. {\bf \textsf{g,}}~Values of the $g$-factor plotted versus their according $\fR$ from panels a (blue) and b (red). We exclusively show the data points for which iso-EDSR is the dominant driving mechanism (yellow regions), {\it i.e.} the gtm-EDSR contribution is $\leq 15\%$ of the measured $\fR$. We fit the data using Eq.~\ref{eq:glso} which assumes iso-EDSR (fit shown by black line with yellow highlight symbolising iso-EDSR), yielding an intrinsic NW $g$-factor of $g_{\rm{NW}} \approx 1$.
}
\label{fig:GFactorRenorm}
\end{figure*}

\noindent{\bf \Large \textsf{Introduction} }\vspace{1pt}\\
Spins in semiconductor quantum dots (QD) have emerged as one of the leading contenders for encoding and processing quantum information \cite{Loss1998, Stano2022, Burkard2023}. Their success is attributed to their competitive coherence times \cite{Laucht2016, Yoneda2017} , the demonstration of robust multi-qubit operations \cite{Hendrickx2020a, Hendrickx2021, Philips2022}, coherent spin control above $1$~K \cite{Yang2020, Petit2020a, Camenzind2022, Huang2024} and their compatibility with industrial fabrication techniques \cite{Zwerver2022, Neyens2023, Weinstein2023}.\\
Among the various semiconductor systems capable of hosting spin qubits, material systems exhibiting strong, intrinsic spin-orbit interactions (SOI), have received increasing attention in recent years \cite{Scappucci2020, Fang2023}. Taking advantage of the SOI, all-electrical spin driving can be implemented via electric dipole spin resonance (EDSR), without the requirement for on-chip micro magnets or microwave antennas. While SOI mediated EDSR allows for compact device architectures \cite{Nowack2007, Watzinger2018}, it has more importantly led to ultrafast Rabi oscillations ranging from $80~$MHz in electrons \cite{Gilbert2023}, to several $100~$MHz \cite{Froning2021, Wang2022} and even up to $1.2$~GHz \cite{Liu2023} in holes, alas, at the expense of coherence. These remarkable gate-speeds have thus raised a pivotal concern for the future of spin qubits with strong SOI \cite{Wang2021}: Do strong couplings to the qubit driving field inevitably lead to increased decoherence due to enhanced couplings with undesired noise sources \cite{Massai2024}?\\ First steps towards reducing the coupling to charge noise have been taken via the modification of global system parameters such as the external magnetic field orientation \cite{Tanttu2019, Piot2022, Hendrickx2024}, however, the fundamental trade-off between speed and coherence has so far prevailed.\\
Here we provide experimental evidence for compromise-free scaling, through the demonstration of a coherence sweet spot that coincides with maximal Rabi driving speeds. Our observations are in agreement with previous theoretical predictions on group IV hole spin qubits \cite{Wang2021, Bosco2021a, Adelsberger2022, Michal2023, Mauro2024}. Additionally, we achieve this Fast and Coherent Tunable Operating Regime (\textit{FACTOR}) all-electrically, by controlling static gate voltages at the individual qubit level. Such local optimizations allow us to respond to the variable electrostatic environments that each qubit experiences, which can be comprised of electric stray fields from neighbouring gates or non-uniform strain.\\
Realising a compromise-free qubit requires navigating the intricate interplay between the tuning parameter, driving mechanism and decoherence channel. Remarkably, certain systems naturally exhibit the conditions for a \textit{FACTOR}, such as hole spins in quasi 1D systems with strong SOI. In these systems, the $g$-tensors often display a high level of anisotropy, which can be notably influenced by electric fields. Consequently, random charge fluctuations couple to the qubit energy, resulting in reduced coherence \cite{Hendrickx2022}. In order to mitigate this issue, configurations with vanishing derivatives of the $g$-factor, with respect to voltage changes, are most promising, as they indicate a reduced coupling of $g$ to charge noise. To realise such a sweet spot, we exploit the properties of a spin qubit hosted inside a squeezed, elongated hole quantum dot, subject to strong SOI, as is naturally provided by the geometry of a Ge/Si core/shell nanowire (NW) \cite{Froning2021}.\\ In such structures, the strong biaxial confinement causes heavy- and light-hole (HH-LH) states to intermix, giving rise to a strong and electrically tunable direct-Rashba spin-orbit interaction (DRSOI) \cite{Kloeffel2011,Kloeffel2018}. At an optimal HH-LH mixture, the spin-orbit strength is expected to reach a maximum as a function of an externally applied electric field. In the presence of strong SOI, when the spin-orbit length $\lso$ becomes comparable to the  dot size $\ldot$, the $g$-factor is reduced \cite{Kloeffel2011, Kloeffel2018, Bosco2021, Froning2021a}. This gives rise to a minimum in $g$ at the point where SOI reaches a maximum. This minimum in $g$ as a function of electric field corresponds to the coherence sweet spot. By additionally employing iso-Zeeman EDSR driving of the qubit \cite{Golovach2006}, the Rabi frequency $\fR$ can be maximised at this very same spot. This ensures an operating regime where the qubit is both fast and long lived without compromising speed or coherence, thus defining our \textit{FACTOR}.\\ \\ \\
\noindent{\bf \Large \textsf{Results} }\vspace{1pt}\\\\
{\bf \large \textsf{Coherent Spin Control at 1.5 K} }\vspace{1pt}\\ We first demonstrate the operation of our Ge/Si NW hole spin qubit at $1.5$~K, adding it to the list of previously demonstrated hot qubits \cite{Yang2020, Petit2020a, Camenzind2022, Huang2024}. A scanning electron micrograph showing a representative device is presented in Fig.~\ref{fig:HotGeSiQubit}a. The device consists of a Ge/Si core/shell NW lying on top of nine bottom gates. For details on the device fabrication we refer to the materials and methods section. By applying positive voltages to the first five bottom gates from the left, the intrinsic hole gas inside the NW is depleted to form a hole double quantum dot (DQD) \cite{Froning2018} with a net-effective hole occupation $(m,n)$ on the left and right dot.
The true total hole occupation has been estimated to be in the range of several dozen \cite{Froning2018, Ungerer2023}, where DRSOI is still predicted to be present due to the large sub band splittings of $\sim 40$~meV \cite{Kloeffel2011,Froning2021a, Bosco2022}.\\ The device is operated at a net-effective charge-transition from~$(1,1)$~to~$(2,0)$ that exhibits Pauli spin blockade (PSB). We apply a positive bias across the NW and measure the current through the DQD, which provides spin to charge conversion due to PSB. This allows to read out the state of the effective spin-$1/2$ system, as shown in Fig.~\ref{fig:HotGeSiQubit}b. The characteristic DC transport signature of PSB can be seen in Fig.~\ref{fig:HotGeSiQubit}c, where the baseline of the bias triangle disappears in the absence of magnetic field (inset \ref{fig:HotGeSiQubit}c).
%
%
%
%
\\
In our setup, the right plunger gate RP, coloured in magenta in Fig.~\ref{fig:HotGeSiQubit}a, is connected to a high frequency line via a bias-tee. This allows for the application of square voltage pulses and microwave bursts to the gate, in addition to DC voltages. The measurements were performed employing a common two-stage pulsing protocol \cite{Maurand2016, Hendrickx2020, Froning2021, Camenzind2022}, as schematically shown in Fig.~\ref{fig:HotGeSiQubit}d.\\
The system is initialized in a spin-blockaded effective (1,1) triplet state,  by employing a specific plunger gate voltage $\VRP$, represented by the magenta stars in Figs.~\ref{fig:HotGeSiQubit}c~and~d. After a short waiting time, the system is pulsed into Coulomb blockade by applying a square voltage pulse, as indicated by the magenta circles in Figs.~\ref{fig:HotGeSiQubit}c and d. While in Coulomb blockade, a microwave (MW) burst of duration $\tburst$ is applied. Pulsing back to the initial voltage of $\VRP$ allows to record a current signal $I$, proportional to the likelihood of a singlet configuration after coherent manipulation. Fig.~\ref{fig:HotGeSiQubit}e shows typical EDSR measurements where the applied MW frequency $\fMW$ is swept against the external magnetic field $\mathbf{B}$, and from which the $g$-factor is extracted. On resonance, the spin is rotated, lifting the spin-blockade which leads to an increased current. By varying the burst duration as a function of magnetic field detuning at a fixed frequency of the microwave drive $\fMW$, coherent Rabi oscillations can be observed as shown in Fig.~\ref{fig:HotGeSiQubit}f, with a gate quality factor of $^{g}Q = \fR \times T_2^{\rm{Rabi}} \approx 5$. These results establish coherent qubit control at $1.5$~K.\\
To gather information about the level of control over the qubit, similar scans to those presented in Figs.~\ref{fig:HotGeSiQubit}e and f were repeated for different electrostatic environments experienced by the qubit whilst remaining within the same charge occupation of the DQD. Each electrostatic configuration of the qubit is defined by the three barrier gate voltages $\VL$, $\VM$ and $\VR$, while the plunger gate voltages $\VLP$ and $\VRP$ were compensated to remain at a fixed readout point.\\ \\
\noindent{\bf \large \textsf{SOI in a Squeezed Quantum Dot}}\vspace{4pt}\\
We characterize each qubit configuration by measuring the Rabi frequency $\fR$ and Land\'e~$g$-factor, $g$, and show their functional dependence on the three individual barrier voltages in Figs.~\ref{fig:GFactorRenorm}a-c. In each of the three studies, the voltages on the other two barrier gates are held at a constant value, indicated by the vertical dashed lines in Figs.~\ref{fig:GFactorRenorm}a-c. The larger responses of $g$ and $\fR$ to the barrier voltages $V_{\rm{L}}$ and $V_{\rm{M}}$, compared to $V_{\rm{R}}$, suggest that the driven qubit is the spin located above the left plunger gate (LP). Moreover, the opposite trends of $g$ and $f_{\rm{R}}$ with respect to voltage are consistent with the theoretical description of an elongated quantum dot in the presence of SOI \cite{Bosco2021}. To understand the observations made in Figs.~\ref{fig:GFactorRenorm}a-c, we first explain how SOI renormalises the $g$-factor, followed by how it can be related to $\fR$ by choosing a specific EDSR driving mechanism.\\ In the case of a quasi 1D, elongated quantum dot, as it is reasonable to assume for our NW, the longitudinal axis of gate-defined confinement is well described by a harmonic potential resulting in a Gaussian envelope of the hole wave function. In the presence of SOI, the spins acquire a helical texture along the NW \cite{Kloeffel2011,Kloeffel2018,Bosco2021,Froning2021a}, provided the external magnetic field $\mathbf{B}$ acting on the g-tensor $\mathbf{\hat{g}}$ gives rise to a Zeeman vector $\mathbf{\hat{g}}\cdot \mathbf{B}$ with a component perpendicular to the SOI axis $\boldsymbol{\alpha}_{\rm{SO}}$. As a result, the hole wave function averages over different spin orientations, leading to a renormalization of the dot $g$-factor (see~Eq.~\ref{eq:glso}~I\;). This effect is especially relevant when the dot extension along the NW, $\ldot$, and spin-orbit length $\lso$, which represents the distance a hole must traverse to undergo a spin rotation due to SOI, are comparable in size. This leads to a minimum in $g$-factor where the SOI is strongest, {\it i.e.} $\lso$ shortest. Relating $g$ to $\fR$, however, additionally requires an assumption on the underlying qubit driving mechanism.
%
%
%
%
\newline We assume that all effects arising from SOI, induced by a change in voltage $\Delta V$,  are captured by a modulation of the $g$-tensor \cite{Crippa2018} while keeping the magnetic field constant, that is $\mathbf{\hat{g}}_0\cdot \mathbf{B} \xrightarrow[]{\Delta V} \mathbf{\hat{g}}_{\Delta V}\cdot \mathbf{B}$. We represent this as a rotation of the Zeeman vector around the perpendicular component of the spin-orbit vector, $\boldsymbol{\alpha^{\perp}}_{\rm{SO}}$. Periodic displacements of the wave function along the NW can result in SOI-mediated Rabi oscillations. We refer to them as iso-Zeeman EDSR (iso-EDSR), if they conserve the modulus of the Zeeman vector $|\mathbf{\hat{g}}\cdot \mathbf{B}|$, Fig.~\ref{fig:GFactorRenorm}d left. In this case $\fR \propto 1/\lso$ \cite{Golovach2006}, and $g$ can be directly related to $\fR$ (see Eq.~\ref{eq:glso}~II\;),
\begin{equation}
\label{eq:glso}
	g~\overset{\rm{I}}{=}~~ g_{\rm{NW}}\cdot e^{-{(\ldot/\lso)}^2} \overset{\rm{II}}{=}g_{\rm{NW}} \cdot e^{-C\cdot\fR^2}\quad.
\end{equation}
Here, $g$ denotes the measured $g$-factor as extracted from Fig.~\ref{fig:HotGeSiQubit}e, $C$ is a fitting parameter, and $g_{\rm{NW}}$ is the confinement-dependent intrinsic NW $g$-factor without renormalization due to SOI, and is assumed constant to first order, thus not capturing residual $g$-modulations from voltage-induced changes of the confinement or other higher order contributions. The derivation is provided in the Supplementary Information. We further note that the functional form of Eq.~\ref{eq:glso} results from the harmonic confinement along the longitudinal axis of the NW and is independent of the microscopic origin of the SOI.\\ As seen in Eq.~\ref{eq:glso}, when reducing $\lso$ (thereby increasing SOI), $g$ is suppressed and $\fR$ increased. This behaviour can be observed in Figs.~\ref{fig:GFactorRenorm}a and b. To obtain a minimum in $g$ and a maximum in $\fR$, as seen in Fig.~\ref{fig:GFactorRenorm}a, the governing SOI has to plateau or show a maximum, as an implicit function of voltage. Ge/Si core-shell NWs as the one used here \cite{Froning2021}, like Ge-hut wires \cite{Watzinger2018} and Si-FinFETs \cite{Camenzind2022}, particularly benefit from DRSOI which is expected to reach a local maximum at moderate electric fields below $10~\rm{MV}/\rm{m}$ \cite{Kloeffel2011,Froning2021,Kloeffel2018,Adelsberger2022}. These predicted electric field ranges are consistent with the voltage range of $\sim100$~mV around the extrema of $g$~and~$\fR$ shown in Fig.~\ref{fig:GFactorRenorm}a, assuming a voltage drop over $\sim50$~nm (gate-pitch). Within the experimental error (see Supplementary Information) the maximum in $\fR$ coincides with the minimum in $g$. The small shift between the extrema can primarily be attributed to residual $g$-modulations, not captured by our model in Eq.~\ref{eq:glso}.\\
In order to facilitate iso-EDSR, displacements of the hole wave function along the NW are desirable, while only minimally varying the dot potential. To this end the MW drive is chosen at gate RP, located as far as possible from the qubit on gate LP. To see where the driving mechanism is consistent with iso-EDSR, we measure the response of $g$ to variations $\Delta\VRP$, to extract $\partial g/ \partial V_{\rm{RP}}$ at a fixed magnetic field $\mathbf{B}$. These responses are shown in Figs.~\ref{fig:GFactorRenorm}e and f, as a function of $\VL$~and~$\VM$.\\ If, while driving, the voltage shifts $\Delta\VRP$ cause significant variations of the target dot potential, the induced Rabi oscillations can be significantly influenced by $g$-tensor modulated EDSR (gtm-EDSR). Such modulations of the $g$-tensor do not conserve the modulus of the Zeeman vector $\mathbf{\hat{g}}\cdot \mathbf{B}$, Fig.~\ref{fig:GFactorRenorm}d right. Since the response $\partial g/ \partial V_{\rm{RP}}$ is computed from the difference in lengths of Zeeman vectors $|\mathbf{\hat{g}}_0 \cdot \mathbf{B}|-|\mathbf{\hat{g}}_{\Delta\VRP} \cdot \mathbf{B}| = \Delta_{|\mathbf{\hat{g}} \cdot \mathbf{B}|}$ under $\Delta\VRP$, we can only provide a rough upper bound to the $g$-tensor modulated Rabi contribution $\fR^{\rm{gtm}}$ which is proportional to a transverse modulation of the Zeeman vector $\Delta_{|\mathbf{\hat{g}} \cdot \mathbf{B}|}^{\rm{gtm}}$. These approximate upper bounds for $\fR^{\rm{gtm}}~\propto~|\Delta_{|\mathbf{\hat{g}} \cdot \mathbf{B}|}^{\rm{gtm}}/\Delta\VRP|~\leq~|\Delta_{|\mathbf{\hat{g}} \cdot \mathbf{B}|}/\Delta\VRP|$ are represented by the bars in Figs.~\ref{fig:GFactorRenorm}a and b. For the calculation of $\fR^{\rm{gtm}}$ and  a qualitative description of the EDSR driving mechanisms we refer to the Supplementary Information.\\
As seen in Figs.~\ref{fig:GFactorRenorm}a, b, e and f, we qualitatively divide the measured data points into two regions, driven primarily by pure iso-EDSR (yellow) and a mix of both, iso- and gtm-EDSR (brown). If $|\partial g/ \partial V_{\rm{RP}}| \rightarrow 0$, also the component attributed to gtm-EDSR, $\fR^{\rm{gtm}} \rightarrow 0$. This leaves iso-EDSR as the only available mechanism that can induce Rabi oscillations.\\
The pairs of $\fR$ and $g$ at specific gate voltages $\VL$ and $\VM$ in Figs.~\ref{fig:GFactorRenorm}a and b, that were classified as pure iso-EDSR (Supplementary Information), are plotted against each other in Fig.~\ref{fig:GFactorRenorm}g, where they are fitted using Eq.~\ref{eq:glso}. From the fit parameter $g_{\rm{NW}} \approx 1$ and the experimental values of $g$, the corresponding $\lso$ were calculated yielding a range from $65$~nm to $150$~nm, assuming $\ldot = 50$~nm (gate-pitch). It is worth noting that this estimation of $\lso$ does not rely on knowledge of the effective mass $\meff$, which can be challenging to estimate in systems with HH-LH mixing, as the one studied here. Further, the calculated values of $\lso$ are in agreement with those obtained via magnetic field spectroscopy \cite{Froning2021a} and qubit measurements in Ref.~\cite{Froning2021} assuming HH-LH mixing.\\\\
\noindent {\bf \large \textsf{Compromise-Free Operation}}\vspace{1pt}
%
%
%
\begin{figure*}[htb]
\centering
\includegraphics[width=1\linewidth]{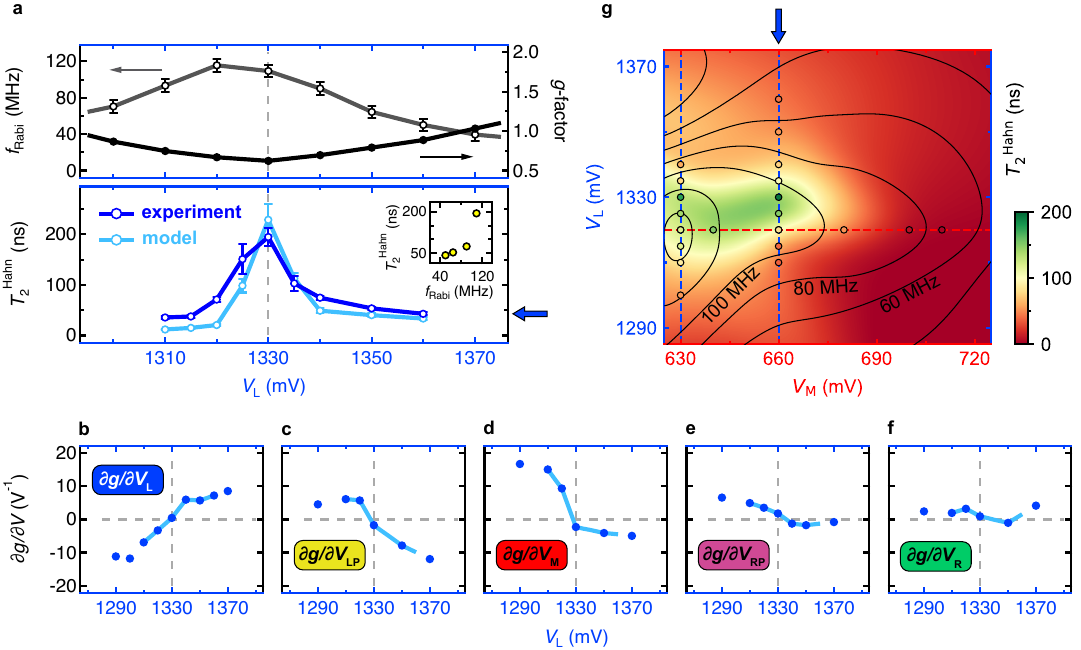} 
\caption{ {\bf \textsf{Compromise-free operation. a,}}~Top: Experimental data of $\fR$ and $g$ as function of $\VL$. Bottom: Dark blue circles show coherence times $T^{\rm{Hahn}}_2$ obtained from fits of the time-dependent readout current measured in Hahn-echo experiments at different voltages $\VL$. Light blue circles show modeled coherence times taking into account $g'_i(\VL)$ from panels b-f using Eq.~\ref{eq:dgdVfit}, yielding $S_{\rm{G}}~=~29~\rm{nV}/\sqrt{\rm{Hz}}$. Vertical dashed line at $V_{\rm{L}} = 1330$~mV shows sweet spot where $T^{\rm{Hahn}}_2$ is maximal. This spot coincides with fastest $\fR$ within experimental errors. Error bars correspond to standard deviations that result from fitting. Inset: Pairs of $\fR$ and $T^{\rm{Hahn}}_2$ from $\VL = 1330-1360$~mV in $10~$mV steps. {\bf \textsf{b-f,}}~Show $g'_i(\VL) := \partial g/ \partial V_{\rm{i}}(V_{\rm{L}})$, as a function of different static voltages on gate $V_{\rm{L}}$. Dark blue data represent values of $V_{\rm{L}}$ for which $g'_i(\VL)$ were measured. Light blue line represents linearly interpolated data needed to match the number of measured data points between $g'_i(\VL)$ and $T^{\rm{Hahn}}_2$, which was fitted in panel a. Vertical dashed line marks sweet spot at $V_{\rm{L}} = 1330$~mV where all five $g'_i(\VL)$ reach near zero, defining the sweet spot. Horizontal dashed line marks the zero-line of the $g'_i(\VL)$. {\bf \textsf{g,}}~The FACTOR: 2D voltage space showing $T^{\rm{Hahn}}_2$ as function of $\VM$ and $\VL$. Filled circles represent measured coherence times $T^{\rm{Hahn}}_2$ extracted analogously to panel a. Data trace corresponding to the dark blue trace on panel a is referenced by the blue arrow on the 2D plot. An additional trace of $T^{\rm{Hahn}}_2$ as a function of $\VL$ for fixed $\VM = 630$~mV is indicated by the vertical dark blue dashed line to the left end of the plot, as well as a trace of $T^{\rm{Hahn}}_2$ as a function of $\VM$ for fixed $\VL = 1320$~mV which is indicated by the horizontal red dashed. Background is obtained by a Gaussian process interpolation serving as a guide to the eye. Black contours show a Gaussian process interpolation of measured Rabi frequencies, highlighting the overlap of maxima in $\THa$ and $\fR$.}
\label{fig:SweetSpot}
\end{figure*} 
Given the extremal behaviour of the g-factor and $\fR$ from Fig.~\ref{fig:GFactorRenorm}a, we measured the Hahn-echo decay times $\THa$ as a function of $\VL$ and compare them in Fig.~\ref{fig:SweetSpot}a. We note that our estimate of $T_2^* \approx 5$~ns was on the order of the duration of the gate pulse duration ranging from $2-8$~ns. These pulse durations can therefore not be assumed to be instantaneous \cite{Cywinski2008}, rendering $T_2^*$ too short to be reliably extracted. We attribute the comparably short $T_2^*$ to the measurement's sensitivity to low-frequency noise, due to the long integration times required by our transport measurements. Details about the dominant noise source in our device are provided in the Supplementary Information.\\ The applied pulse sequence involved a $\pi_x/2$ pulse, a refocusing $\pi$ pulse, and a $\pi_{\varphi}/2$ pulse with a sinusoidally varying phase as a function of the free evolution time $\tau$, allowing for a more robust fit of the Hahn decay. Assuming $1/f^{\beta}$ noise, the exponent of the Hahn-echo decay $\alpha = 1+\beta$ was defined as a global fit parameter, shared among all data sets, and converged to $\alpha \approx 1$. This value for $\alpha$ was extracted for the frequency range sampled by the Hahn-echo experiment from approximately $1/\tau_{\rm{max}} = 12.5$~MHz up to $\fR$, where $\tau_{\rm{max}}$ is the longest waiting time between pulses in the Hahn experiment. Given the operation temperature of $1.5$~K, a white noise spectrum could be expected, as observed in other spin qubit experiments \cite{Camenzind2022, Petit2018, Huang2024, Liles2024}. Above $1.7$~K, the qubit readout in this device significantly degrades. As seen in Fig.~\ref{fig:SweetSpot}a, the coherence peaks at a gate voltage $\VL$ where the g-factor is minimal and $\fR$ is maximal, within the experimental error margin. This observation clearly indicates a compromise-free scaling regime in which the fastest qubit operation times coincide with the coherence sweet spot. As seen in the inset of Fig.~\ref{fig:SweetSpot}a the {\textit FACTOR} is characterized by a simultaneous fourfold increase in coherence and threefold increase in operation speed. Defining $Q^{\rm{Hahn}} = \fR \cdot \THa$ as a quantitative measure for quality, this improvement corresponds to a more than tenfold increase in performance, from $\sim 2$ to $\sim 21$. Further, the qualitative signature of the sweet spot remains robust regardless of the choice of noise color $\beta$, as shown in the Supplementary Information.
\\ Next, we analyze the response of $g$ to small voltage fluctuations of the gate voltages $\Vi$ as a function of $\VL$, denoted as $\partial g/\partial \Vi(\VL)$. We refer to $\partial g/\partial \Vi(\VL)$ as $g'_i(\VL)$ for brevity. Figs.~\ref{fig:SweetSpot}b-f show the $g'_i(\VL)$ of the five gates used in our experiments. One striking commonality among all five $g'_i(\VL)$ is the near-zero point crossing at the sweet spot voltage $\VL^* = 1330$~mV, indicating minimal modulations of the Zeeman vector magnitude $|\mathbf{\hat{g}} \cdot \mathbf{B}|$. Slight variations of the readout point between experimental runs and residual modulations of $g_{\rm{NW}}$ can lead to minor offsets in the zero-point crossing and residual decoherence \cite{Bosco2021,Bosco2021a}.
\\We suspect the noise to mainly originate from charge traps in the self-terminated, native $\rm{SiO}_2$ layer, uniformly covering the NW shell, see Supplementary Information. This brings a source of noise very close to the $\rm{Ge}$ core, where the hole wave function resides. We model the noise as voltage fluctuations on all the gates with one common noise spectrum, characterized by a spectral density $S_{\rm{G}}$ and exponent $\beta$. Considering the appropriate noise filter function for a Hahn-echo and $\beta = 0$ (as determined before), the characteristic decay rate can be expressed as \cite{Piot2022},
\begin{equation}
	\label{eq:dgdVfit}
	\frac{1}{\THa} = 2\pi^2~\left( \sum_{i} \left( \frac{\partial g}{\partial \Vi}~S_{\rm{G}} \right)^2  \right) \qquad.
\end{equation}
The best fit to the measured $\THa$, employing Eq.~\ref{eq:dgdVfit}, is obtained with a white noise spectral density of $S_{\rm{G}}(f_0/f) =  29~\rm{nV}/\sqrt{\rm{Hz}}$. In keeping with Ref.~\cite{Piot2022} we choose  $f_0 = 1/\tau_{\rm{max}} = 12.5$~MHz, corresponding to the frequency above which the noise was probed. This is an approximate value set by the duration of our shortest Hahn-experiments, and provides an estimate of the noise frequencies our Hahn-experiment is susceptible to. The modeled coherence times are shown in Fig.~\ref{fig:SweetSpot}a. No additional lever arms are required in Eq.~\ref{eq:dgdVfit}, as they are already captured by the $g'_i(\VL)$. Lastly, we repeat similar coherence measurements for different values of $\VM$ and show in Fig.~\ref{fig:SweetSpot}g that the \textit{FACTOR} is not limited to a point in voltage space. In fact, the existence of a "sweet ridge" can be observed, which shows a weaker dependence of the coherence time $\THa$ on $\VM$ as compared to $\VL$, over comparable voltage ranges. As shown by the black contours in Fig.~\ref{fig:SweetSpot}g, the  Rabi frequency $\fR$ follows the trend of $\THa$, reaching frequencies above $100$~MHz surrounding the coherent region in green. The experimental data points for the contour plot of $\fR$ are shown in the Supplementary Information.\\\\
\noindent {\bf \Large \textsf{Discussion}}\vspace{1pt}\\
We report on the existence of a \textit{FACTOR} for which the coherence sweet spot coincides with the fastest qubit operation speeds. We demonstrate that our \textit{FACTOR} originates from the renormalisation of the $g$-factor yielding a minimum in $g$ and thus $g'_i(\VL) = 0$, when the SOI is maximised for a hole quantum dot subject to strong, quasi 1D confinement. Further, maximal Rabi frequencies are achieved for iso-EDSR driving of the qubit. We favour this driving mechanism by applying the MW drive to a remote gate, and identify static gate voltage ranges for which the qubit is indeed dominantly driven by iso-EDSR. Our experimental observations thus overturn the conventional wisdom that fast qubit operations impose a toll on qubit life times.\\ 
While the presented data were recorded on a single device, experiments on devices fabricated with NWs from the same growth batch \cite{ConesaBoj2017,Froning2018} suggest broad gate tunability and low levels of disorder, reducing the likelihood that the observed behaviour arises from an anomalous disorder effect. However, due to the manual NW deposition-method used in the fabrication of our devices, some additional strain may be applied to the NWs. This could induce slight offsets in the position of the extremum of the SOI with respect to gate-voltage. While homogenous strain would lead to offsets in the chemical potential, inhomogeneous strain could influence the direction and amplitude of the effective DC-electric field applied across the dot which breaks the inversion symmetry to induce SOI.\\ Nevertheless, such static electric fields can be readily compensated for {\it in situ}, at the individual qubit level, via corresponding gate voltage adjustments, making our {\it FACTOR} independent of any global experimental parameters. We thus expect additional strain to not affect the results qualitatively. Our observed non-monotonous behaviour of a Rashba-type SOI of holes confined to quasi 1D, remains best explained by the electric-field tunable HH-LH mixing described within the DRSOI framework \cite{Kloeffel2011}. The required electric field ranges are consistent with our experiments \cite{Kloeffel2018, Bosco2021, Wang2021}.\\
Furthermore, we have established coherent control of a hole spin qubit in a Ge/Si NW at $1.5$~K, with qubit operation speeds and coherence times on par with previous experiments performed at mK temperatures \cite{Froning2021}. This achievement renders our platform compatible with on-chip classical control electronics \cite{Vandersypen2017}. While better qubit performance can be expected at lower T \cite{Dial2013}, other device aspects may also play a crucial role. In particular, the proximity of noisy interfaces such as a native $\rm{SiO}_2$ layer \cite{Donnelly2024}, which in our device is $\sim 2.5$~nm away from the qubit. However, since our demonstrated mechanism changes the qubit's susceptibility to the noise, we expect our observations to change quantitatively but not qualitatively, when changing the operating temperature or the distance of noise sources relative to the qubit.\\
The transferability of our concept to other spin qubit platforms is described by a set of
key criteria:
{\bf i)} The principal mode of qubit control is intrinsic or synthetic SOI. Dephasing is dominated by electric noise fluctuations coupling to the spin through SOI. {\bf ii)}~A non-monotonic SOI tuning response as a function of the noise parameter is essential. In an elongated dot, when $\lso \approx \ldot$, the effective g-factor is renormalized. When SOI is maximal (minimal $\lso$) this yields a minium in g with reduced voltage sensitivity and minimal decoherence. {\bf iii)}~Under iso-Zeeman driving, this same point corresponds to a maximum in $\fR$, creating an optimal operation point where speed and coherence are simultaneously maximal. {\bf iv)}~For hole systems, the non-monotonic SOI response relies on controlling the mixing of HH and LH states. This is achieved when the transversal confinement in both axes becomes comparable in strength and aspect ratio, and is significantly stronger than the longitudinal confinement.\\
Within this framework, the question may be raised, whether the conditions leading to the observed compromise-free operation might be translated to 2D hole spin qubit platforms. Thanks to their relaxed lithographic constraints resulting from the comparatively low in-plane effective hole masses \cite{Scappucci2020}, lateral squeezing gate electrodes could be used to induce strong, quasi 1D confinement \cite{Bosco2021} {\color{black} to planar Ge systems such as the one presented in, {\it e.g.}, \cite{Hendrickx2020}, where a $\sim 16$~nm thick quantum well is buried $\sim 22$~nm below the surface. To define the lateral squeezing axis, we suggest the fabrication of parallel finger gates (squeezing gates) separated by $\sim20$~nm, on the order of the thickness of the quantum well.
Systems with significantly deeper wells and effective masses comparable to that of the HH (light in-plane effective mass) \cite{Camenzind2019} have achieved similar confinement lengths. These remain feasible even for in-plane effective masses approaching that of the LH \cite{Vandersypen2017,Mills2019,Langrock2023}. A second gate layer oriented perpendicular to the squeezing axis could then be used to divide the longitudinal axis into $\sim 50-100$~nm segments, which should be on the order of $\lso$. A driving gate that is offset along the longitudinal confinement axis would favor iso-Zeeman driving.}\\
One may pose the question whether compromise-free operation can be transferred to electron-based systems. Conventional Rashba SOI could potentially also be tuned by electric fields to reach a maximum in SOI, but might require larger electric fields than the ones used here, as the SOI is primarily governed by the fundamental band gap of the semiconductor obtained in the third order of a multi-band perturbation theory \cite{Winkler2003, Carballido2021}. For electron systems using synthetic SOI, we cautiously note that nanomagnet arrays could induce g-factor renormalization if engineered to produce a spatial winding of the magnetic field on the order of the spread of the wavefunction. Motivated by recent nanomagnet structures \cite{Aldeghi2023}, we propose designing magnetic field gradients that are maximal on the longitudinal axis of the elongated quantum dot and roll-off perpendicular to the axis of elongation.
Again, using lateral squeezing gates, the QD could be positioned on the extremum of the magnetic field gradient, yielding a similarly maximal artificial SOI as a function of lateral electric field, making fluctuations of the confinement voltage perpendicular to the elongated dot axis less detrimental. While the transferability of compromise-free operation may be more straight-forward for holes, it is more speculative for electron-based platforms whose coherence is not dominantly limited by spin–orbit-induced dephasing.
Should nanomagnet-based architectures however prevail, strategic positioning of the quantum dots relative to the magnetic field gradients may become an important consideration to prevent coherence from being compromised by the stronger couplings associated with synthetic SOI.\\
Finally, it is important for either electrons or holes that the SOI is strong enough to enter the strong SOI regime, where $\lso~\approx~\ldot$ holds.
By demonstrating the feasibility of a \textit{FACTOR} in a hole spin qubit, our work offers a new angle from which to approach fault-tolerant quantum computation without sacrificing high qubit operation speeds.\\\\
\noindent {\bf \Large \textsf{Methods}}\vspace{1pt}\\
\small\noindent {\bf \textsf{Device Fabrication.}} The QD device featured a set of nine bottom gates, each with a width of $\sim 20$~nm and a pitch of $\sim 50$~nm. The gates were fabricated on an intrinsic Si $(100)$ chip with $290$~nm of thermal $\rm{SiO}_2$ using electron beam lithography (EBL). After cold development, the bottom gates were metallized with $1$~nm/$9$~nm of Ti/Pd, respectively. To provide electrical insulation between the bottom gates and the NW, $175$ cycles ($\sim 20$~nm) of $\rm{Al}_2\rm{O}_3$ were grown by atomic layer deposition at approximately $225^\circ$~C using atomic layer deposition. In an effort to improve the quality of the gate dielectric, the chip underwent annealing in a $20$~mbar forming gas atmosphere ($\rm{N}_2~92\%,~\rm{H}_2~8\%$) for 15 minutes at $300^\circ$~C, prior to NW deposition. Details on the impact of the annealing process on gate hysteresis and qubit coherence are described in the Supplementary Information.\\
A single Ge/Si core/shell NW was deterministically placed in a perpendicular orientation to the nine bottom gates. The NW has a core radius of $\sim 10$~nm and a shell thickness of $\sim 2.5$~nm. The exact in-plane angle, however, remains unknown. Subsequently, ohmic contacts were patterned by EBL and metallized with Ti/Pd layers of $0.3$~nm/$50$~nm of Ti/Pd, respectively, following a $10$ second dip in buffered hydrofluoric acid to locally remove the native $\rm{SiO}_2$ layer in the defined contact region. Fig.~\ref{fig:HotGeSiQubit}a presents a scanning electron micrograph of an analogously fabricated device from the same batch, representative of the measured device.\\
\noindent {\bf \textsf{Measurement Apparatus.}} The experimental setup featured a variable temperature insert (VTI) in a liquid helium bath with the sample mounted below the $1$K pot (base temperature $1.5$~K). The VTI was equipped with a solenoid magnet controlled by an Oxford Instruments IPS magnet power supply. DC voltages were supplied by a Basel Precision Instruments digital-to-analog converter (LNHR~$927$) and filtered on a dedicated filter PCB (second-order RC low-pass filter, cutoff frequency $8$~kHz). Fast gate pulses and IQ control pulses were generated on a Tektronix AWG $5204$. A Rohde \& Schwarz SGS100A Vector Signal Generator was used to generate the qubit control pulses through IQ modulation. The gate- and control pulses were combined using a Wainwright WDKX11 diplexer and delivered to the sample PCB using attenuated coaxial lines. A bias-tee on the sample PCB was used to combine the high frequency pulses with a DC bias. The DC current through the NW was amplified by a Basel Precision Instruments current-to-voltage converter (LSK389A) with a gain of $10^9$ and measured using a National Instruments DAQ card (USB-6363). The Vector Signal Generator output was pulse-modulated by a Zurich~Instruments MFLI lock-in amplifier at $77.777$~Hz to enhance the signal-to-noise ratio of the qubit measurements.\\ 
\noindent {\bf \textsf{Data Analysis.}}
The $g$-factors were measured as described in Fig.~\ref{fig:HotGeSiQubit}e, and were extracted for each considered electrostatic configuration defined by the barrier voltages $\VL$, $\VM$ and $\VR$. The positions of the resonance condition, with respect to $\fMW$ at fixed $\bf{B}$, are obtained by fitting each column to a Gaussian. The slope of a linear fit to the center positions of the Gaussians then yields the $g$-factor. The Rabi frequencies $\fR$ were extracted from fits to $I(t) = I_{\rm{offset}}+I_0\sin(2\pi\fR\tburst+~\phi_0)\exp(-\tburst/T_2^{\rm{Rabi}})$. Here, $I_{\rm{offset}}$ is an offset, $I_0$ the amplitude, $\phi_0$ a phase shift and $T_2^{\rm{Rabi}}$ the characteristic decay time. The characteristic decay times of each individual Hahn-echo experiment $\THa$ were obtained from a global fit of all echo-experiments using $I(t)~=~I_{\rm{offset}}~+~I_0\sin(2\pi f_{\varphi}\tau_{\rm{wait}}~+~\phi_0)\exp(-(\tau/\THa)^{\alpha})$, and one shared parameter $\alpha = \beta + 1$. Furthermore $f_{\varphi}$ describes the frequency at which the phase of the pulse $\pi_{\varphi}/2$ was artificially varied as a function of the free evolution time $\tau$.
\\
\noindent {\bf \textsf{Measurement Details.}}
The derivatives $\partial g/\partial \Vi(V_j) := g'_{i}(V_j)$ presented in Fig.~\ref{fig:SweetSpot}b-f were extracted in three different ways at fixed $\fMW$: {\bf  i) $\boldsymbol{ g'_{i}(\VL)}$, for $\boldsymbol i\in\{\rm{LP, M ,R}\} $ :} The readout point was defined by fixing all gate voltages. The derivatives were then obtained by recording the $g$-factors at manually varied voltages $\Delta\Vi = 2-4$~mV, without losing readout. {\bf  ii)~$\boldsymbol{ g'_{\rm{RP}}~(V_j)}$,~for~$\boldsymbol j\in\{\rm{L, M}\} $:} For a fixed readout point, the depth $\Delta V_{\rm{CP}}$, by which the system was pulsed into Coulomb blockade, was varied by up to $10$~mV while all other voltages were held constant. Only the DC voltage on $\VRP$ was algorithmically adjusted for each $\Delta V_{\rm{CP}}$ by a linear correction factor in order to keep the readout point fixed. As in i), the derivatives were computed by fitting the slope of the recorded $g$-factor versus the variation of $\Delta V_{\rm{CP}}$. {\bf iii)~$\boldsymbol{ g'_{\rm{L}}(\VL)}$ :} We first computed the derivative of the recorded $g(\VL)$ presented in Fig.~\ref{fig:GFactorRenorm}a with respect to $\VL$, yielding $g'_{\rm{L}}(\tilde{V}_{\rm{L}})$. Here, $\tilde{V}_{\rm{L}}$ is used to highlight that for each value of $\VL$, the voltage $\VLP$ was compensated as otherwise the readout point would have been lost over the considered range of $\VL$ due to the large cross-capacitance. Therefore, in order to obtain the true derivative $g'_{\rm{L}}(\VL)$, the influence of $\VLP$ was subtracted to first order via $g'_{\rm{L}}(\VL) = g'_{\rm{L}}(\tilde{V}_{\rm{L}}) - g'_{\rm{LP}}(\VL)\cdot\Delta\VLP/\Delta\VL$. Here, $g'_{\rm{LP}}(\VL)$ is taken from i) and the compensation $\Delta\VLP/\Delta\VL \approx -0.49$.\\
All qubit measurements were performed at a microwave frequency of $2.79$~GHz, to accommodate the specifications of the diplexer and to avoid resonances of the RF-wiring, requiring the qubit to be operated at fields between $\sim 100-400$~mT, given range of $g$-factors. The power of the microwave frequency was fixed at $P_{\rm{MW}} = -13.4$~dBm power and $V_{\rm{IQ}} = 300$~mV IQ voltage amplitude, corresponding to an AC excitation of $V_{\rm{ac}} = 7.8$~mV at the driving gate RP. The Rabi chevron shown in Fig.~\ref{fig:HotGeSiQubit}f was taken near the optimal operating regime at $\VL=1320$~mV, $\VM=660$~mV, $\VR=1020$~mV. The EDSR resonance shown  in Fig.~\ref{fig:HotGeSiQubit}e was recorded at a fixed $\tburst=4$~ns.\\ The Hahn-echo experiments were taken at $2$~seconds integration time and depending on amplitude of the transport current, up to 50 averages were taken of each trace to improve the signal to noise ratio.
\\\\
\noindent {\bf \Large \textsf{Data Availability}}\vspace{1pt}\\
\noindent The data supporting the plots of this paper are available at the Zenodo repository at https://doi.org/10.5281/zenodo.10223162.
\\\\
\noindent {\bf \Large \textsf{Acknowledgments}}\vspace{1pt}\\
\noindent We thank Y.-M. Niquet, S. Geyer and F.N.M.~Froning for the fruitful discussions. This work was supported by the NCCR SPIN, the Swiss Nanoscience Institute (SNI), the Georg H. Endress Foundation, the Swiss National Science Foundation (SNSF), the EU H2020 European Microkelvin Platform (EMP) project (Grant No.~$824109$) and the Topologically Protected and Scalable Quantum Bits (TOPSQUAD) project (Grant No.~$862046$). J.C.E. acknowledges support from CNPq/Brazil (Grant No 301595/2022). L.C.C. acknowledges support from a Swiss NSF mobility fellowship (P$2$BSP$2-$200127).\\ \\
\noindent {\bf \Large \textsf{Author Contributions}}\vspace{1pt}\\
\noindent M.J.C., S.S., T.P. and D.M.Z. conceived of the project and planned the experiments. M.J.C., S.S. and P.C.K. fabricated the device. A.L. and E.P.A.M.B. synthesized the NWs. S.B. developed the theoretical framework for squeezed quantum dots with inputs from D.L.. R.M.K. studied the gate dielectric quality. M.J.C. and T.P. designed the experimental setup with inputs from L.C.C.. M.J.C., S.S. and T.P. executed the experiments. J.S. contributed to the extended data sets supervised by N.A.. M.J.C., R.S.E., T.P. and S.B. developed the requirements for a \textit{FACTOR} and analyzed the data with input from J.C.E. and D.M.Z.. D.M.Z. supervised the project. M.J.C.~wrote the manuscript with inputs from all the authors.\\ \\
\noindent {\bf \Large \textsf{Competing Interests}}\vspace{1pt}\\
\noindent The authors declare no competing interests.
%
%
%
%
%
%

%

%
%
\end{document}


\title{Supplementary Information: Compromise-Free Scaling of Qubit Speed and Coherence}

\author{Miguel J. Carballido}
\email{miguel.carballido@unibas.ch}
\affiliation{Department of Physics, University of Basel, Klingelbergstrasse 82, CH-4056 Basel, Switzerland}

\author{Simon Svab}
\affiliation{Department of Physics, University of Basel, Klingelbergstrasse 82, CH-4056 Basel, Switzerland}

\author{Rafael S. Eggli}
\affiliation{Department of Physics, University of Basel, Klingelbergstrasse 82, CH-4056 Basel, Switzerland}

\author{Taras~Patlatiuk}
\affiliation{Department of Physics, University of Basel, Klingelbergstrasse 82, CH-4056 Basel, Switzerland}

\author{Pierre~Chevalier~Kwon}
\affiliation{Department of Physics, University of Basel, Klingelbergstrasse 82, CH-4056 Basel, Switzerland}

\author{Jonas~Schuff }
\affiliation{Department of Materials, University of Oxford, Oxford OX1 3PH, United Kingdom}

\author{Rahel~M.~Kaiser}
\affiliation{Department of Physics, University of Basel, Klingelbergstrasse 82, CH-4056 Basel, Switzerland}

\author{Leon C. Camenzind}\thanks{Currently at: CEMS, RIKEN, Wako, Saitama 351-0198, Japan}
\affiliation{Department of Physics, University of Basel, Klingelbergstrasse 82, CH-4056 Basel, Switzerland}

\author{Ang~Li}\thanks{Currently at: Institute of Microstructure and Properties of Advanced materials, Beijing University of Technology, Beijing, 100124, China}
\affiliation{Department of Applied Physics, TU Eindhoven, Den Dolech 2, 5612 AZ Eindhoven, The Netherlands}

\author{Natalia Ares}
\affiliation{Department of Engineering Science, University of Oxford, Oxford OX1 3PJ, United Kingdom}

\author{Erik~P.A.M~Bakkers}
\affiliation{Department of Applied Physics, TU Eindhoven, Den Dolech 2, 5612 AZ Eindhoven, The Netherlands}

\author{Stefano~Bosco}\thanks{Currently at: QuTech and Kavli Institute of Nanoscience, Delft University of Technology, Delft, The Netherlands}
\affiliation{Department of Physics, University of Basel, Klingelbergstrasse 82, CH-4056 Basel, Switzerland}

\author{J. Carlos Egues}
\affiliation{Department of Physics, University of Basel, Klingelbergstrasse 82, CH-4056 Basel, Switzerland}
\affiliation{Instituto de F\'isica de S\~{a}o Carlos, Universidade de S\~{a}o Paulo, 13560-970 S\~{a}o Carlos, S\~{a}o Paulo, Brazil}

\author{Daniel Loss}
\affiliation{Department of Physics, University of Basel, Klingelbergstrasse 82, CH-4056 Basel, Switzerland}
\affiliation{CEMS, RIKEN, Wako, Saitama 351-0198, Japan}

\author{Dominik M. Zumb\"uhl}
\email{dominik.zumbuhl@unibas.ch}
\affiliation{Department of Physics, University of Basel, Klingelbergstrasse 82, CH-4056 Basel, Switzerland}

\date{August 15, 2025}
\maketitle

\onecolumngrid

\tableofcontents

\newpage
\section{\texorpdfstring{$g$}{TEXT}-Factor Renormalization due to SOI under iso-Zeeman Driving}
\begin{figure}[htb!]
\centering
\includegraphics[width=0.49\linewidth]{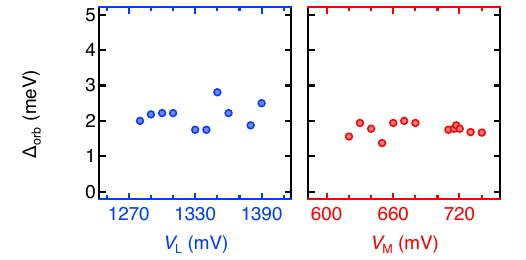} \caption{{\bf \textsf{ Estimation of $\boldsymbol \dorb$.}}~Extracted from the ratio between the separation from the bias triangle base line to the first excited state, divided by the triangle height at $V_{\rm{SD}}=5$~mV, as a function of $\VL$ (left) and $\VM$ (right).}
\label{fig:dorb}
\end{figure}

\begin{spacing}{1.5}
\noindent Assuming a harmonic confinement potential along the longitudinal axis of the NW, and in the presence of SOI, the measured $g$-factor is renormalized by a Gaussian envelope function \cite{Kloeffel2013,Bosco2021},
\begin{equation}
\label{eq:gSO}
	g = g_{\rm{NW}}~\exp{-\left(\frac{\ldot}{\lso}\right)^2}\qquad,
\end{equation}
where $g_{\rm{NW}}$ is the intrinsic $g$-factor derived from the microscopic confinement of the NW, $\ldot$ is the dot size along the direction of lowest confinement and $\lso$ is the spin-orbit length defined here as the distance a hole has to traverse along the NW to have its spin flipped due to SOI.\\
The application of an oscillating electric field to gate RP in the presence of SOI, gives rise to an oscillating effective magnetic field $\Beff(t)$, with magnitude \cite{Golovach2006}, 
\begin{equation}
\label{eq:Beff}
	\Beff(t) = 2B~\frac{\ldot^2}{\lso}~\frac{eE_{\rm{MW}}(t)}{\dorb} \qquad,
\end{equation}
where $e$ is the elementary charge, $E_{\rm{MW}}$ is the a.c.~electric field in the dot generated by the microwaves, $\dorb = \hbar^2\ldot^{-2}\meff^{-1}$ is the orbital level splitting with $\meff$ the effective hole mass. The effective magnetic field $\mathbf \Beff(t)$, drives the Rabi oscillations, at the Rabi frequency $\fR = g~\mub|{\mathbf \Beff}|/2h$, with $g$ parallel to $\mathbf B$ and h Planck's constant.\\
To stay on resonance, the variation of $g$, induced by changes of the electrostatic environment, is compensated with a proportional change of $\mathbf B$, to match the Larmor frequency set by a fixed MW frequency $h\fMW = g\mub|\mathbf B|$. This effectively makes $\fR$ independent of $g$,
\begin{equation}
\label{eq:fR}
	\fR = \frac{\fL}{\lso}~\frac{\ldot^2~e|E_{\rm{MW}}(t)|}{\dorb} \qquad.
\end{equation}
Combining Eqs.~\ref{eq:gSO} and \ref{eq:fR}, yields the relation between $g$ and $\fR$,
\begin{equation}
\label{eq:gfR}
	g = g_{\rm{NW}}~\exp{-\left(C\cdot\fR^2\right)}\qquad,
\end{equation}
with the fitting constant $C = \dorb^3~\meff~\fMW^{-2}~\hbar^{-2}~e^{-2}~|E_{\rm{MW}}|^{-2}$. Fig.~\ref{fig:dorb} shows the estimated orbital spacings $\dorb$ as a function of $\VL$ and $\VM$ to be roughly constant over the operated voltage range. Here $|E_{\rm{MW}}|$ is constant as the frequency and power of the microwave signal were held constant for all experiments.
\end{spacing}
\newpage
\clearpage
\section{Non-Monotonicity of the Qubit Parameters}
\begin{figure}[htb!]
\centering
\includegraphics[width=1\linewidth]{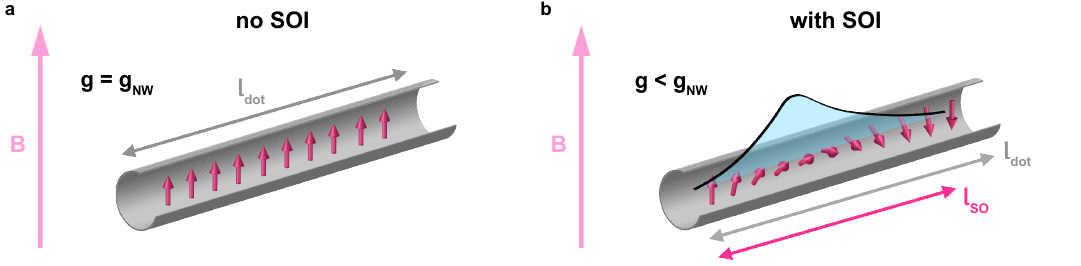} \caption{{\bf \textsf{Schematic depiction of the spin texture with and without SOI in quasi-1D.~a,}}~In the presence of a magnetic field {\bf B} and absence of SOI, the effective $g$-factor is constant along the NW.~{\bf \textsf{b,}}~In the presence of SOI, the envelope function of the dot averages over the helical spin texture leading to a reduced effective $g$-factor.}
\label{fig:lso_avg}
\end{figure}

\begin{spacing}{1.5}
\noindent Because the argument in the exponential suppression of $g_{\rm{NW}}$ solely depends on $\fR\propto\lso^{-1}$, any non-monotonous dependence of the SOI on the gate voltage, represented through $\lso(V)$, is inherited by $\fR$ and indirectly by the effective $g$-factor through the Gaussian term written in Eq.~1 (Main). To provide a more intuitive picture for the latter, in the presence of a magnetic field $\mathbf B$ and from the rest-frame of the quantum dot in the absence of SOI, the envelope function of the dot sees the same spin-polarization along the entire NW, {\it i.e.} $g = g_{\rm{NW}}$, as schematically shown in Fig.~\ref{fig:lso_avg}a. When the two lengths $\lso\approx\ldot$ become comparable, the envelope function averages over the helical spin texture leading to an effective $g$-factor that is suppressed, as schematically shown in Fig.~\ref{fig:lso_avg}b.\\\\
The non-monotonic SOI, which both the $g$-factor and $\fR$ depend on, can be explained as follows:\\
At zero electric field, the structural inversion symmetry of the NW is preserved, leading to a vanishing SOI. When an electric field $E_x$ is applied perpendicular to a NW with a cylindrical cross-section \cite{Kloeffel2018}, the inversion symmetry is broken, resulting in an increase of the spin-orbit energy $E_{\rm{SO}} \propto E^2_z$. As the electric field is further increased, the degree of heavy-hole (HH) and light-hole (LH) mixing gradually changes, until at (much) higher fields, the system begins to resemble a 2D confinement. In this regime, the ground state becomes predominantly HH-like \cite{Scappucci2020}, and the linear-in-k direct-Rashba spin-orbit interaction (DRSOI) is weakened due to the large HH-LH splitting $\Delta_{\rm{HH-LH}}$ exceeding 100 meV \cite{Terrazos2021}, which governs the strength of the SOI $\propto \Delta^{-1}_{\rm{HH-LH}}$ \cite{Winkler2003}.\\
In Ge/Si core-shell NWs like the one used in our work, the DRSOI is expected to reach a local maximum at moderate electric fields below 10 MV/m \cite{Adelsberger2022,Kloeffel2011,Kloeffel2018}. As mentioned in the main text, these predicted electric field ranges correspond to the voltage range of approximately $100$~mV observed around the extrema of $g$ and $\fR$ in our Fig. 2a, assuming a voltage drop across a gate pitch of about $50$~nm.\\
While it is difficult to provide a direct ``ultimate proof'' of the mechanism, our experimental results are consistent with the theoretical framework providing a most plausible explanation for the observed effect, based on the interplay between electric field strength and HH-LH mixing.
\end{spacing}

\clearpage
\newpage
\section{Experimental Error on \texorpdfstring{$\fR$}{TEXT}}
\begin{figure}[htb!]
\centering
\includegraphics[width=1\linewidth]{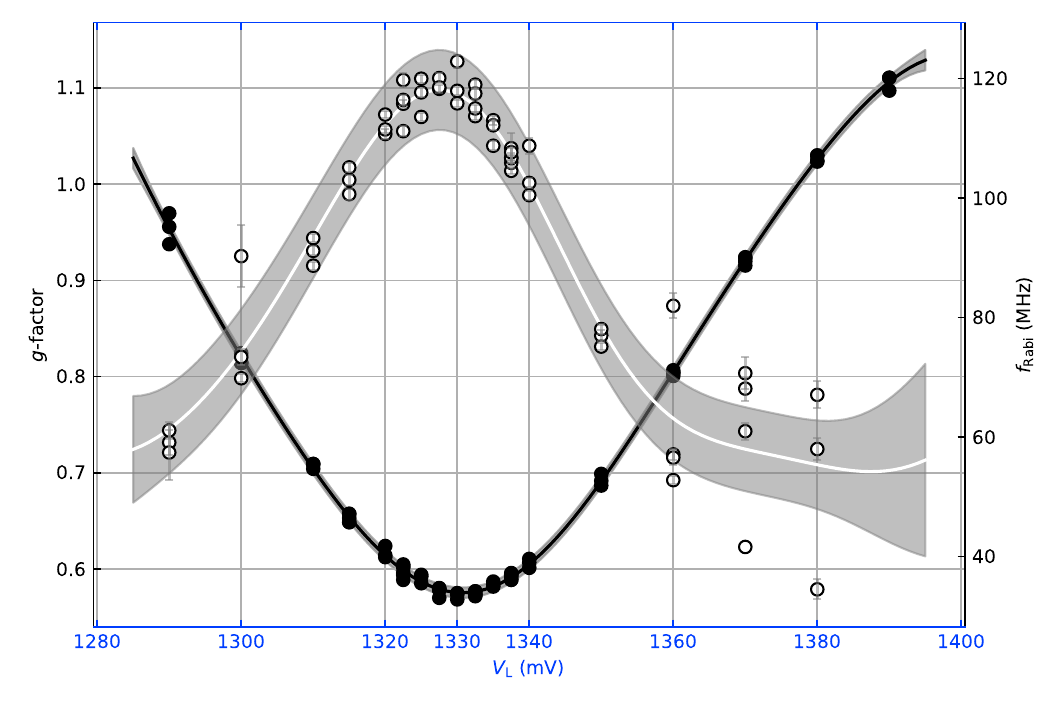} \caption{{\bf \textsf{ Reproducibility of the extrema in $\boldsymbol{\mathbf \fR}$ and g.}}~Experimental data of the qubit g-factor (solid black circles) and Rabi frequency $f_{\rm{Rabi}}$ (open circles), as a function of the left barrier gate voltage $\VL$, at fixed gate voltages $\VM = 660$~mV and $\VR = 1020$~mV. The fit uncertainty of the measured Rabi oscillations is illustrated by grey error bars. For both the $g$-factor and $\fR$, Gaussian process fits are applied using a radial basis function kernel. The solid lines denote the average prediction of the Gaussian processes, while the shaded bands show the uncertainty of the Gaussian processes corresponding to one standard deviation.}
\label{fig:Revisitgf}
\end{figure}
%
\begin{spacing}{1.5}
\noindent The data series presented in Fig.~\ref{fig:Revisitgf} was collected in multiple sequences from low to high $V_\text{L}$. The extremal points of $\fR$ and $g$ coincide within $5$~mV considering the experimental variation between runs. For $\fR$ this statistical variation corresponds to $\sim10$~MHz around the maximum (approx. $10\%$), while the variation of the $g$ is negligibly small. \\ The observed spread of the measured quantities between runs can be attributed to slight variations of the readout point, in turn slightly varying $g$ and $\fR$. Furthermore, charge switchers which shift the readout point relative to the bias triangle can influence $g$ and $\fR$ as well. We note that the traces presented in Fig.~\ref{fig:Revisitgf} were taken at a far later point in time relative to the first measured data points which were presented in the main section. Additionally the sample has experienced  several thermal cycles from $1.5$~K to $9$~K. Taking into account both  these circumstances and the reproducibility of the data, speaks for the stability of the device and the observed effects. To more accurately explain differences in the extremal points of $\fR$ and $g$ with regards to their positioning in gate voltage $\VL$, as well as residual decoherence, a more detailed analytical model must be considered that captures the voltage dependences of the intrinsic NW $g$-factor, $g_{\rm{NW}}$ \cite{Bosco2021,Bosco2021a}, which in this work was assumed to be constant. We would like to direct special attention to Fig.~8b in Ref.~\cite{Bosco2021a}.
\end{spacing}
\newpage
\section{Possible Manifestations of g-Tensor Modulation vs Iso-Zeeman Driving on the Measured Zeeman Vector}
\label{sec:drivemechs}
\begin{figure}[htb!]
\centering
\includegraphics[width=1\linewidth]{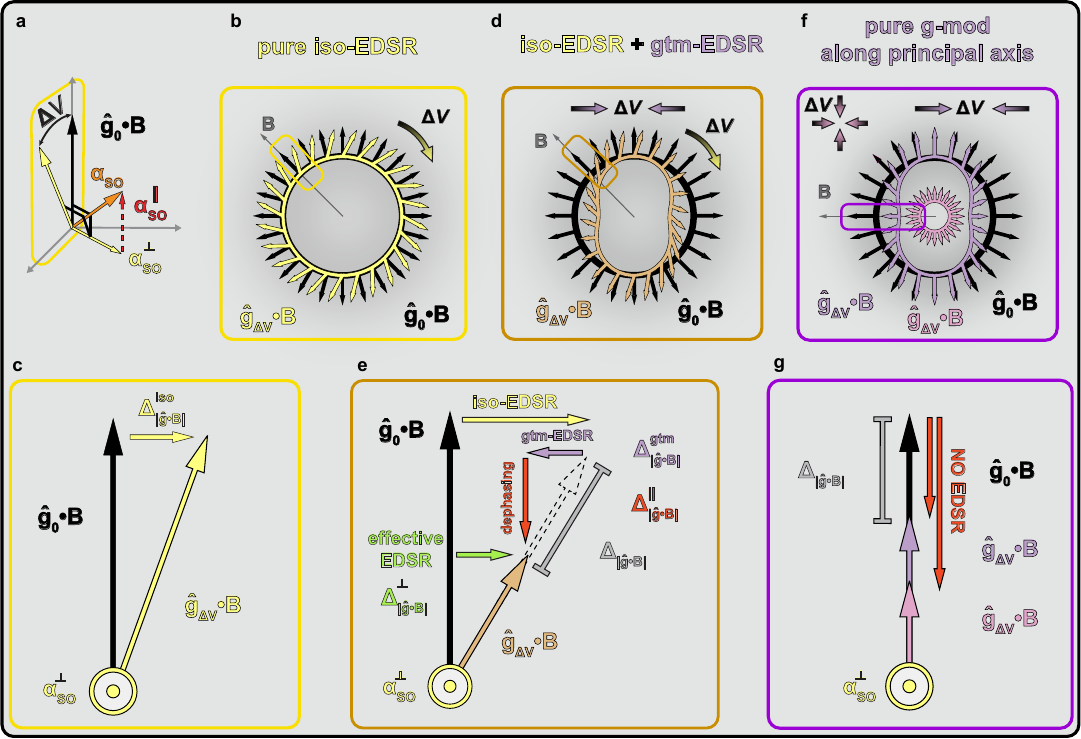} \caption{{\bf \textsf{Possible manifestations of qubit driving mechanisms on the measured Zeeman vector. a,}}~Decomposition of the effective spin-orbit vector $\aso$, into perpendicular $\asope$ and parallel $\asopa$ components relative to $\mathbf{\hat{g}}_0\cdot\mathbf{B}$. {\bf \textsf{b,}}~Shown in black is the Zeeman vector field of an arbitrary, here isotropic, $g$-tensor $\hat{g}_0$. Each vector represents a specific direction of magnetic field $\mathbf B$ with constant magnitude. Yellow shows the Zeeman vector field in the case of iso-EDSR, for which the magnitude of an arbitrary Zeeman vector is conserved, and the vector is solely rotated proportional to the strength of SOI and applied amplitude $\Delta V$. No distortions of the $g$-tensor and therefore Zeeman vector field are present. {\bf \textsf{c,}} Blown up image of one specific direction of $B$ as indicated in grey in panel b. To first order, the only component acting on the incident Zeeman vector is perpendicular (small angle between $\mathbf{\hat{g}}_0\cdot \mathbf{B}$ and $\mathbf{\hat{g}}_{\rm{\Delta V}}\cdot \mathbf{B}$. {\bf \textsf{d,}}~Zeeman vector field showing the manifestation of both, iso-EDSR and gtm-EDSR (brown). {\bf \textsf{e,}} The length of the Zeeman vector is not conserved. The perpendicular component to the drive originating from modulations of the $g$-tensor (violet), in this specific example, counteracts the pure iso-EDSR drive (yellow) resulting in an effective EDSR which induces the Rabi oscillations (green). {\bf \textsf{f-g,}}~Edge cases when the $g$-tensor modulations occur along the principal magnetic axes (purple) or the dot is symmetrically compressed (or expanded) in the form of a "breathing"-mode (pink), which do not give rise to EDSR. }
\label{fig:drivemechs}
\end{figure}
\begin{spacing}{1.5}
\noindent Provided a constant magnetic field $\bf B$, we capture all effects arising from SOI induced by periodic voltage displacements $\Delta V$, by changes of the $g$-tensor \cite{Crippa2018}.
Formally, that is $\mathbf{\hat{g}}_0\cdot \mathbf{B}\xrightarrow[]{\Delta V} \mathbf{\hat{g}}_{\Delta V}\cdot \mathbf{B}$. Further, we consider solely the vectorial component of the effective spin-orbit vector $\aso$ that is perpendicular to $\mathbf{\hat{g}}_0 \cdot \mathbf{B}$, hence $\asope$. Mechanisms involving the parallel component of $\aso$ with respect to $\mathbf{\hat{g}}_0 \cdot \mathbf{B}$, namely $\asopa$, and are responsible for e.g. the longitudinal driving of the spin through the application of an off-resonant microwave tone \cite{Bosco2022,Bosco2023}, are not considered here. A vectorial decomposition of $\aso$ is shown in Fig.~\ref{fig:drivemechs}a.\\
 In the case of SOI mediated Rabi oscillations, we can make a distinction between two EDSR driving mechanisms, based on the way they manifest on the magnitude of the resonant Zeeman vector. When the dot is subject to an infinitesimal change in voltage $\Delta V$ in the presence of SOI, the action of the induced effective magnetic field $\BSO$ around which the spin precesses, can be represented by a rotation of the Zeeman vector $\mathbf{\hat{g}}_0 \cdot \mathbf{B}$ around $\asope$.\\
Assuming no further anisotropies of the $g$-tensor are introduced by $\Delta V$, we define the induced rotations of the spin that arise while the modulus of the Zeeman vector is conserved, as iso-Zeeman EDSR (iso-EDSR), Fig.~\ref{fig:drivemechs}b. To first order, for a small enough variation $\Delta V$, and thus small angle between $\mathbf{\hat{g}}_0 \cdot \mathbf{B}$ and $\mathbf{\hat{g}_{\Delta V}\cdot B}$, the only component acting on the incident Zeeman vector $\mathbf{\hat{g}}_0 \cdot \mathbf{B}$ is perpendicular, $\Delta^{\rm{iso}}_{|\mathbf{\hat{g}} \cdot \mathbf{B}|}$, Figs.~\ref{fig:drivemechs}c. As outlined in \cite{Golovach2006}, iso-EDSR is expected to be the dominant qubit driving mechanism, whenever the motion of the dot induced by periodic voltage shifts $\Delta V$ does not come along significant variations of the dot potential.\\
Next, we classify contributions to EDSR that do not conserve the modulus of $\mathbf{\hat{g}}_0 \cdot \mathbf{B}$ as $g$-tensor modulated EDSR (gtm-EDSR). It arises whenever changes in the dot potential modulate the magnetic axes of the $g$-tensor and as a result anisotropically modify the Zeeman vector field, Fig.~\ref{fig:drivemechs}d. This mechanism introduces both, perpendicular $\Delta^{\rm{gtm}}_{|\mathbf{\hat{g}} \cdot \mathbf{B}|}$ and parallel $\Delta^{\parallel}_{|\mathbf{\hat{g}} \cdot \mathbf{B}|}$ components to the incident Zeeman vector, Fig.~\ref{fig:drivemechs}e. Usually, iso-EDSR is accompanied by gtm-EDSR, due to the fact that voltage changes on the gates which displace the dot likely also affect the potential landscape. In such cases, gtm-EDSR can counter act (or enhance) the driving component expected from pure iso-EDSR, yielding an effective EDSR drive $\Delta^{\perp}_{|\mathbf{\hat{g}} \cdot \mathbf{B}|} = \Delta^{\rm{iso}}_{|\mathbf{\hat{g}} \cdot \mathbf{B}|} \pm \Delta^{\rm{gtm}}_{|\mathbf{\hat{g}} \cdot \mathbf{B}|}$. It is not directly possible to derive the driving strength caused by gtm-EDSR from only measuring $\Delta_{|\mathbf{\hat{g}} \cdot \mathbf{B}|}$, as it is the case in our experiments. However, this absolute change in magnitude of the Zeeman vector does provide an upper bound on the $g$-tensor modulated contribution to the Rabi oscillations, $\fR^{\rm{gtm}}\propto\Delta^{\rm{gtm}}_{|\mathbf{\hat{g}} \cdot \mathbf{B}|}$.\\
Finally, there remain some corner cases of highly symmetric modulations of the dot potential ("breathing modes"), or specific orientations of $\mathbf{B}$ which affect solely the principal magnetic axis longitudinally, and can be considered as pure $g$-tensor modulation, Fig.~\ref{fig:drivemechs}f. These scenarios however do not lead to Rabi oscillations, as they lack components perpendicular to the incident Zeeman vector and therefore do not induce EDSR, Fig.~\ref{fig:drivemechs}g.\\\\
The simplest Hamiltonian to describe pure iso-EDSR in our system can be obtained following Ref.~\cite{Trif2008}, where an expression is derived to describe the precession of a spin induced by the effective magnetic field that arises due to SOI.\\
We use the coordinate system  introduced in Fig.~1 of the main text. Our Hamiltonian consists of a kinetic term, a harmonic confinement term along the y-axis (longitudinal axis of the NW) corresponding to the axis of weakest confinement, a Zeeman term with magnetic field $\boldsymbol{B} = (B_x,0,0)$, an oscillating electric field along the $y$-direction $\boldsymbol{E_{\rm{ac}}}~=~E_0\cos(\omega_{\rm{ac}}t)$, and a spin orbit term with spin-orbit strength $\alpha^{\perp}_{\rm{SO}}$ which is perpendicular to $\boldsymbol{B}$ and the longitudinal axis of the NW, yielding:
\begin{equation*}
	H^{\rm{iso}}_{\rm{EDSR}} = \frac{\hbar^2\hat{k}^2_y}{2m_{\rm{eff}}}~+~\frac{1}{2}m_{\rm{eff}}\omega^2_0y^2~+~\frac{1}{2}g_{\rm{NW}}\mu_{\rm{B}}B_x\hat{\sigma}_x~+~eE_0\cos(\omega_{\rm{ac}}t)y~+~\alpha^{\perp}_{\rm{SO}}\hat{\sigma}_z\hat{k}_y \quad,
\end{equation*}
where $\hbar$ is the reduced Planck Constant, $m_{\rm{eff}}$ is the effective electron mass, $\hat{k}_y$ is the 1D momentum operator along the $y$-direction, $\omega_0$ is a constant for the parabolic confinement potential, $\omega_{\rm{ac}}$ is the frequency of the applied ac-field, $g_{\rm{NW}}$ is the intrinsic NW $g$-factor, $\mu_{\rm{B}}$ is the Bohr magneton, $\hat{\sigma}_i$ are the spin-1/2 Pauli operators and $e$ is the elementary charge.\\
With the momentum of the dot set along $\hat{k}_y$, the SOI Hamiltonian would more generally read $\alpha_{\rm{SO}}~\boldsymbol{n}~\cdot~\boldsymbol{\hat{\sigma}}~\hat{k}_y$. Here, the parameter representing the spin-orbit strength $\alpha_{\rm{SO}}$ is inversely proportional to the subband splitting between heavy (HH) and light (LH) holes $\alpha_{\rm{SO}} \propto \Delta^{-1}_{\rm{HH-LH}} \approx (20~{\rm{mV}})^{-1}$ and is obtained by a down-projection from the Luttinger-Kohn Hamiltonian onto the lowest energy HH-LH subspace \cite{Kloeffel2011,Lu2005}.\\
The normal vector $\boldsymbol{n}$ selects the direction of the spin-orbit vector, and depends non-trivially on the static electric field $\boldsymbol{E}$ that is used to break the inversion symmetry as well as strain. Relative to the magnetic field $\boldsymbol{B}$, one can define a component of the spin-orbit vector $\boldsymbol{\alpha_{\rm{SO}}} = \alpha_{\rm{SO}}~\boldsymbol{n}$ that is perpendicular, $\boldsymbol{B}\perp \boldsymbol{\alpha^{\perp}_{\rm{SO}}}$, and one that is parallel, $\boldsymbol{B}\parallel \boldsymbol{\alpha^{\parallel}_{\rm{SO}}}$. When driving at resonance, $\omega_{\rm{ac}} = \omega_{\rm{L}}$, the component that induces Rabi oscillations is $\boldsymbol{\alpha^{\perp}_{\rm{SO}}}$ (where  $|\boldsymbol{\alpha^{\perp}_{\rm{SO}}}| = \alpha^{\perp}_{\rm{SO}}$). In the Hamiltonian above, we chose this component to act along the z-direction, perpendicular to both the magnetic field and the longitudinal axis of the NW.\\
In an ideal setup, we expect the maximal value of $\alpha_{\rm{SO}}$ to enter the Hamiltonian. However, without a vector magnet to precisely calibrate the magnetic field direction, it is unlikely to perfectly reach the conditions of $\boldsymbol{B}\perp \boldsymbol{\alpha_{\rm{SO}}}$ or $\boldsymbol{B}\parallel \boldsymbol{\alpha_{\rm{SO}}}$. Thus, for angles in between, there is likely always a component of $\boldsymbol{\alpha_{\rm{SO}}}$ that will be perpendicular to $\boldsymbol{B}$ around which the spin can precess. The component of the spin-orbit vector $\boldsymbol{\alpha_{\rm{SO}}}$ that points parallel to $\boldsymbol{B}$, {\it i.e.} $\boldsymbol{\alpha^{\parallel}_{\rm{SO}}}$, has no effect on the spin unless a far off-resonant microwave tone is applied to induce longitudinal driving of the spin \cite{Bosco2023}. 
\end{spacing}
\newpage
\section{Impact of the Gate Dielectric Quality on Qubit Coherence}
\begin{figure}[htb!]
\centering
\includegraphics[width=1\linewidth]{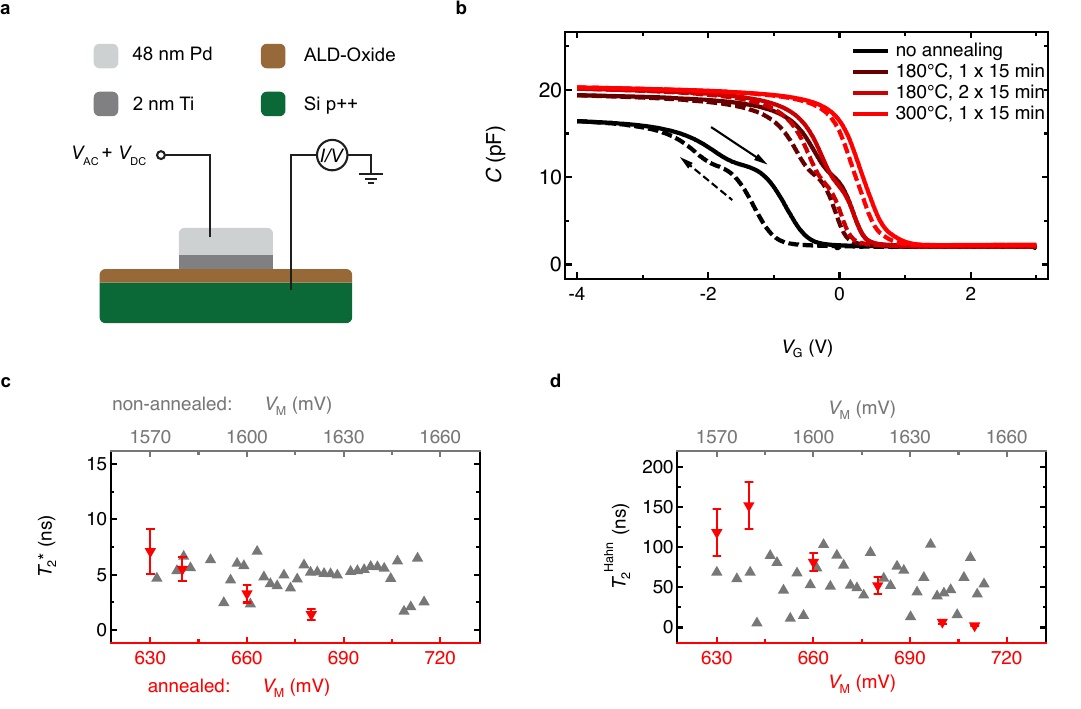} \caption{{\bf \textsf{Impact of Gate Dielectric Quality on Qubit Coherence.~a,}}~Schematic of the MOSCAP used for CV profiling. The device consists of a heavily p-doped Si-substrate covered by $\sim24$~nm of ALD- oxide grown on top of the native Si-oxide of the substrate. The top electrode is defined by an $80~\mu\rm{m}\times80~\mu$m Ti/Pd square. {\bf \textsf{b,}}~CV-curves of a MOSCAP taken as a reference prior to (black), and after subsequent annealing runs. The temperatures and durations are indicated in the legend. The sweep directions are indicated by the arrows. {\bf \textsf{c,}}~Measured values of the Ramsey free evolution time $T_2^*$ as a function of gate voltage $\VM$ for the non-annealed device (grey) and the annealed device (red). {\bf \textsf{d,}}~ Same as {\textsf{c}} but measuring the free evolution time of a Hahn-Echo experiment $\THa$. Error bars correspond to standard deviations that result from fitting.}
\label{fig:OxiCo}
\end{figure}

\begin{spacing}{1.5}
\noindent Couplings to charge noise originating from impure oxides close to the quantum dot, in which the spin qubit is embedded, pose a limit for coherence \cite{Yoneda2017, Culcer2009}.
For the Ge/Si NW devices discussed in this work, the closest suspects are the native Silicon oxide on the shell of the nanowire and the ALD-grown $\rm{Al}_2\rm{O}_3$ oxide used as gate dielectric. Here, we investigate the impact of ALD-grown gate dielectric quality on the qubit coherence of our devices.\\
It has been reported that annealing of oxides can remove trapped charges at the oxide-semiconductor interface \cite{Schroder2005}.
 We thus compare whether an annealed ALD-grown oxide layer shows any improvement of qubit coherence times as compared to an untreated oxide. To qualitatively assess the dielectric quality, we perform capacitance-voltage (CV) profiling experiments on test-metal-oxide-semiconductor-capacitor devices (MOSCAP), as described  in Fig.~\ref{fig:OxiCo}a. The MOSCAPs consist of a material stack resembling that of the qubit devices, but with the difference of using a less resistive heavily p-doped Si substrate to facilitate the capacitance measurements. A DC voltage $V_{\rm{G}}$ from a DAC (Basel Precision Instruments SP927) is combined with an AC voltage $V_{\rm{AC}}$ generated by a lock-in amplifier (Stanford Research Systems SR830) and connected to the metal contact of a MOSCAP with a needle prober. The other needle is placed on a region with the exposed p-doped Si substrate and connected to an IV converter (Basel Precision Instruments SP983c), which is then fed back to the lock-in amplifier. The capacitance is calculated from:
\begin{equation}
\label{eq:cvspec}
	C = \frac{{I}_{AC}^Y}{2\pi f V_{AC}} \qquad,
\end{equation}
where ${I}_{AC}^Y$ is the measured out-of-phase component of the current signal, and $f$ the lock-in frequency. Upon measuring the CV-curves of the untreated MOSCAP devices, they are annealed in a rapid thermal annealing oven (MBE-Komponenten GmbH AO500). The CV-curves of the devices annealed at different temperatures are shown in Fig.~\ref{fig:OxiCo}b for increasing duration or temperature.
The hysteretic behavior of the curves observed in Fig.~\ref{fig:OxiCo}b can be attributed to undesired interface charge traps between the semiconductor and the oxide \cite{Massai2024}. This feature is used as a crude indicator for the oxide quality. Subsequent annealing runs significantly reduce the level of the observed hysteresis and are thus associated with an improvement of the dielectric quality. We note that temperatures beyond $300^{\circ}$~C were not tested to prevent damage to the Ti/Pd bottom gates of the devices. From $400^{\circ}$~C onwards, the bottom gates would show strong signs of deformation.\\
In summary, we attribute the reduction of hysteresis observed in the CV-curves taken after a $300^{\circ}$~C annealing step, to a qualitative improvement of the interface between the p-type substrate and the $\rm{Al}_2\rm{O}_3$ oxide. A more thorough analysis would require to also test the effect of annealing on undoped substrate along with performing measurements at cryogenic temperatures. A further limitation is posed by directly probing the chip surface rather than fabricating an ohmic contact to the doped Si substrate, leading to gate voltage offsets due to the needle-semiconductor contact-interface. 
To investigate any effects of the annealed ALD oxide on qubit coherence, we compare ${T_2}^*$ and $\THa$ measured in the annealed device (described in the main text) with the values in an unannealed device, whose measured data are overlaid in Fig.~\ref{fig:OxiCo}c and \ref{fig:OxiCo}d. We note that for the given bias triangle in the unannealed qubit device, no dependence of the qubit coherence on the center gate voltage was observed. Further, no clear difference in coherence times can be observed between both devices, indicating that annealing of the ALD oxide layer did not noticeably improve the qubit quality.\\ We therefore suspect the main source of charge noise to be the native Silicon oxide on the nanowire shell, and our coherence to be dominantly limited by $T_2$ processes and not $T_1$ processes (see comment further below). From a materials perspective, first steps towards improved coherence may focus on improving the quality of interfaces and surfaces \cite{Sangwan2024} by, for example, terminating the Si-shell with high quality thermal oxides within the growth chamber, as opposed to native oxides that grow in an uncontrolled environment when the material is exposed to ambient air.\\
{\bf Additional Comment on ${\mathbf T_1}$:} As measured in similar NW devices \cite{Hu2011}, $T_1$ is expected to be {\it ca.}~$200-500~\mu$s for the magnetic fields we operate our qubits at, which is orders of magnitude longer than the recorded $T_2$ times.\\ It is nevertheless important to think about scenarios where the coherence may become $T_1$-limited. For example, when the spin qubit were to be strongly coupled to a non-ideal resonator, whereby the losses of the cavity to the photon bath will impact $T_1$ of the spin qubit (Purcell effect).\\
In the occasion that the qubit were indeed limited by $T_1$-processes, there still is a remedy to counteract this problem by going to higher magnetic fields. The gate speed (here represented by the Rabi frequency $\fR$) scales linearly with the spin-orbit coupling strength $\alpha_{\rm{SO}}$ and linearly with the Larmor frequency $\omega_{\rm{L}}$, and hence the magnetic field $B$, such that $f_{\rm{Rabi}} \propto \omega_{\rm{L}}~\alpha_{\rm{SO}} \propto B ~\alpha_{\rm{SO}}$. Furthermore, the $T_1$-decay rate $\Gamma_1 = 1/T_1$ (which we wish to be small) scales quadratically with $\alpha_{\rm{SO}}$ but inverse proportionally with $\omega_{\rm{L}}$, such that $\Gamma_1 \propto \alpha^2_{\rm{SO}}/\omega_{\rm{L}} \propto B^2/\omega_{\rm{L}}$. The ratio of the two yields an expression $f_{\rm{Rabi}}/\Gamma_1 \propto \omega_{\rm{L}}^2/\alpha_{\rm{SO}} \propto B^2/\alpha_{\rm{SO}}$. As can be seen, the detrimental effect of stronger SOI could be compensated by operating the qubit at slightly higher magnetic fields.
\end{spacing}
\clearpage
\newpage
\section{Estimation of \texorpdfstring{$g$}{TEXT}-Tensor Modulated Contribution to \texorpdfstring{$\fR$}{TEXT}}
\begin{figure}[htb!]
\centering
\includegraphics[width=0.49\linewidth]{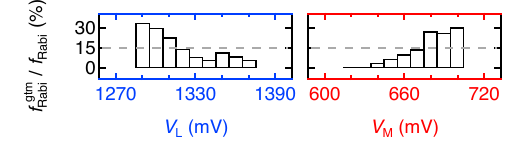} \caption{{\bf \textsf{Ratio of $\boldsymbol \fR^{\rm{\bf gtm}} \boldsymbol /\boldsymbol\fR$.}}~Calculated contribution to $\fR$ coming from gtm-EDSR represented as the bars. The dashed line corresponds to $15\%$ which was used as our threshold for classification.}
\label{fig:percent}
\end{figure}

\begin{spacing}{1.5}
\noindent To first order a periodic voltage fluctuation $d\VRP$ on the driving gate RP, will give rise to a gtm-EDSR contribution of magnitude \cite{Michal2023},
\begin{equation}
	\label{eq:gtEDSR}
	\fR^{gtm}(V) = \frac{\eta}{2}\;\frac{\partial g(V)}{\partial \VRP}\; \frac{\mub}{h}\;B\;V_{\rm{MW}} \qquad,
\end{equation}
where $V_{\rm{MW}}$ is the amplitude of the microwave voltage applied to gate RP, $\partial g/\partial \VRP$ is the response of the $g$-factor to a variation $\VRP$, $\mub$ Bohr's magneton, $h$ Planck's constant, $B$ is the magnetic field amplitude at the qubit resonance and $\eta$ is an efficiency accounting for the distance from the driving gate to the assumed qubit location over gate LP. The applied microwave voltage arriving at the sample is $V_{\rm{MW}} = 7.8$~mV and was obtained from a power calibration. To estimate the efficiency $\eta$, we estimate the horizontal and vertical distance of the expected qubit location relative to the driving gate, $d_{\rm{hor}} = 100$~nm and $d_{\rm{vert}} = 30$~nm (accounting for $\sim 20$~nm of dielectric and the NW radius of $10$~nm). This roughly results in $\eta \approx 30\%$.\\
As explained in Sec.~\ref{sec:drivemechs} $\partial g/\partial V_i$ provides solely an upper bound to the perpendicular component relevant for the $g$-modulated drive. Based on this estimate, datapoints with contributions to the measured Rabi frequencies above $\fR^{\rm{gtm}}/\fR \geq 15\%$ of the measured value of $\fR$ are considered significantly affected by $g$-tensor modulation and were excluded from the fit in Fig.~{\color{blue} 2}g of the main text. The calculated percentages of $\fR^{\rm{gtm}}/\fR$ are shown in Fig.~\ref{fig:percent}.
\end{spacing}
\newpage
\section{Effect of \texorpdfstring{$\beta$}{TEXT} on Computed and Measured Decoherence Rates}
\begin{spacing}{1.5}
\noindent The best fit of the Hahn-decays was obtained for $\beta=0$. We further note that the qualitative feature of the coherence sweet spot is robust with respect to the choice of $\beta$ as seen in Figs.~\ref{fig:THe}a-d. We provide an exemplary Hahn-echo data set in Fig.~\ref{fig:HahnExemp} below. The point at $\VL = 1330$~mV shows some stronger dependence on the choice of noise exponent as opposed to the other points which we explain in the following. Overall, the nature of transport measurements performed in this work, sets an upper bound to the free evolution time $\tau$ of the Hahn measurements. If each hole loaded into the double dot were to be measured with efficiency $\eta = 1$, we would expect the maximum possible current $e/t_{\rm{cycle}}$, where $e$ is the elementary charge and $t_{\rm{cycle}}=2\times(t_{\pi}+2t_{\pi/2}+2\tau)$, which yields $e/t_{\rm{cycle}} \approx 1$~pA for $t_{\rm{cycle}}\approx150$~ns. The leakage currents of our experiments however correspond to efficiencies on the order of $\eta_{\rm{meas}} \approx 0.1$. This inefficiency can be attributed to co-tunneling events, the inefficient loading of the double dot due to the elevated temperature of operation of $1.5$~K or random variations of the device behaviour after a thermal cycle of the device. In the case of $\VL = 1330$~mV, the highest $\tau$ used did not suffice to observe a significant characteristic decay of the Hahn-echo. Further increasing of $\tau$ would have lead to transport currents which are too small to measure. Therefore in this case, the fit was more sensitive on the choice of $\beta$.\\For the models of decoherence rates for non-zero $\beta$, we use the generic expression as derived in the Supplementary Information of \cite{Piot2022}:
\end{spacing}
\begin{equation}
	\label{eq:genericPiot}
	\frac{1}{\THa} = 2\pi~\left(C_{\beta}f_0^{\beta} \sum_{i} \left( \frac{\partial g}{\partial \Vi}~S_{\rm{G}} \right)^2  \right)^{\frac{1}{\beta+1}} \qquad,
\end{equation}
\begin{spacing}{1.5}
    
with $C_{\beta} = 2\sin(\frac{\beta\pi}{2})(2^{1-\beta}-1)\Gamma(-1-\beta) $, $\Gamma$, the Gamma function $\Gamma$, and $S_{\rm{G}} = S_{\rm{G}}(f_0/f)$ is the noise spectral density at reference frequency $f_0 \approx 12.5$~MHz for our measurements.\\ \\
{\bf Details of the Fit Proceedure:} To be more precise, one may also write $T_2^{\rm{Hahn}}(V_{\rm{L}})$ and $\frac{\partial g}{\partial V_i}(V_{\rm{L}})$ to denote that these are both functions of $V_{\rm{L}}$. We hereby assumed one global value for the noise spectrum $S_{\rm{G}}$ ($S_{\rm{G},i} = S_{\rm{G}}$) weighting each of the $\frac{\partial g}{\partial V_i}(V_{\rm{L}})$ equally, as we believe the main source of noise to be the native $\rm{SiO}_2$ shell homogeneously covering the NW. The position and hence the susceptibility to the noise background is then again reflected by the range of amplitude of each of the curves $\frac{\partial g}{\partial V_i}(V_{\rm{L}})$ in panels 3b-f in the main text, and corresponds to the location of the qubit relative to the gates, {\it i.e.} the farther away the gate, the smaller the range of values of $\frac{\partial g}{\partial V_i}(V_{\rm{L}})$. The least squared fit we perform takes into account all values of $V_{\rm{L}}$ and the fit coefficient provided is the one which simultaneously minimized the differences between fitted and the experimentally measured values of $T_2^{\rm{Hahn}}(V_{\rm{L}})$ for their specific values of $V_{\rm{L}}$. This means that all of the five curves and each of their data points in $\frac{\partial g}{\partial V_i}(V_{\rm{L}})$ were taken into account with their corresponding values of $V_{\rm{L}}$ to perform the fit and we did not pick a single specific value $V_{\rm{L}}=V^*_{\rm{L}}$ at a specific gate-voltage.
\end{spacing}

\begin{figure}[htb!]
\centering
\includegraphics[width=1\linewidth]{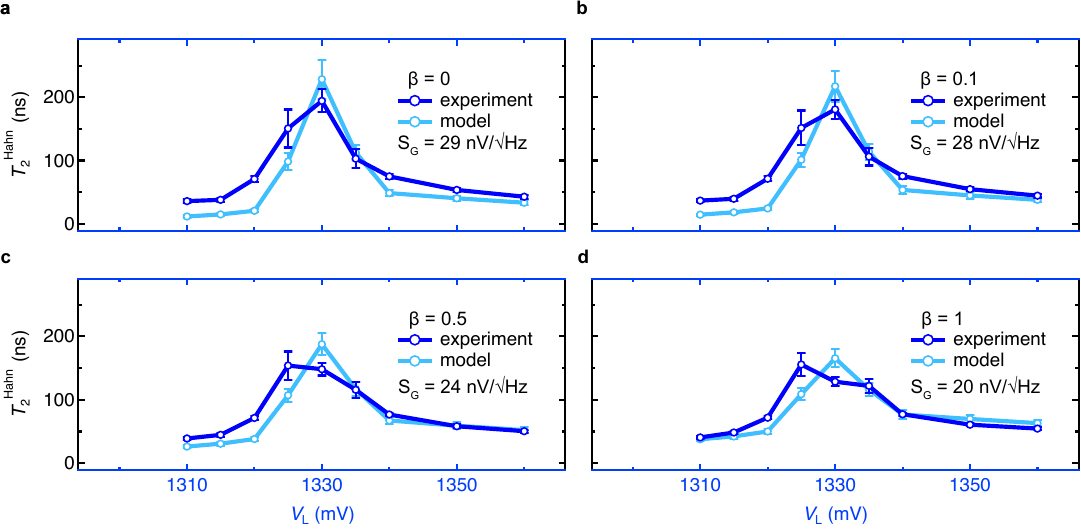} \caption{{\bf \textsf{Analysis of the experimental Hahn-echo data and model for $\boldsymbol{\beta}\in\{0,~0.1,~0.5,~1\}$.~a,}}~Identical traces as presented in Fig.~3a of the main text for $\beta = 0$. {\bf \textsf{b-d,}}~Traces for $\beta \in \{0.1,~0.5,~1\}$. All models for non-zero $\beta$ were computed using Eq.~\ref{eq:genericPiot}. Error bars correspond to standard deviations that result from fitting.} 
\label{fig:THe}
\end{figure}
\begin{figure}[htb!]
\centering
\includegraphics[width=1\linewidth]{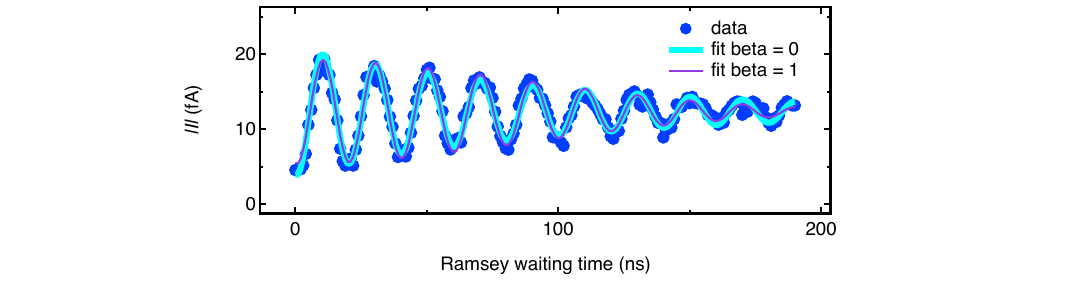} \caption{ {\bf \textsf{Exemplary data set of a Hahn-echo experiment at $\boldsymbol{V_{\rm{L}} = 1335}$~mV.}}~The qubit state is initialised along the quantization axis ($z$-axis).The pulse sequence applied to the qubit consists of a $\pi_x/2$-pulse to bring the initial state into the equatorial plane. This is followed by a waiting time $\tau$ after which a $\pi_x$-pulse is applied to refocus the state during a waiting time of $\tau$. Finally a second $\pi_x/2$-pulse is applied in order to measure the qubit state along the quantization axis. The data points in blue show the current amplitude as a function of waiting time $\tau$ between the $\pi/2$-pulse and the $\pi$-refocussing pulse. In keeping with the Supplementary Information S6 we show the data points of the Hahn-echo experiment with both $\beta = 0$ and $\beta = 1$, yielding $T_{2,\beta=0}^{\rm{Hahn}} = 103\pm15$~ns and $T_{2,\beta=1}^{\rm{Hahn}} = 122\pm11$~ns, respectively. The frequency at which the phase of the second pulse $\pi_{\varphi}/2$ varied was set to $f_{\varphi} = 50$~MHz. The errors correspond to standard deviations that result from fitting.} 
\label{fig:HahnExemp}
\end{figure}

\newpage
\clearpage
\section{2D Voltage Maps of \texorpdfstring{$\fR$}{TEXT} and \texorpdfstring{$\THa$}{TEXT}}
\begin{figure}[htb!]
\centering
\includegraphics[width=1\linewidth]{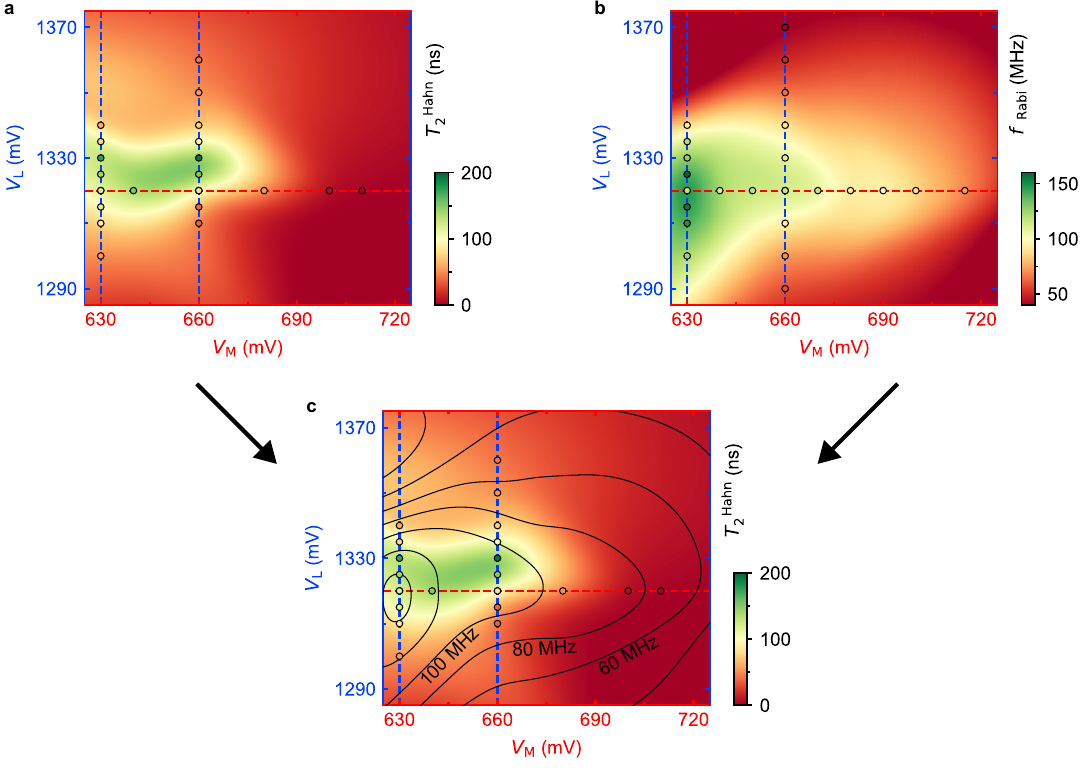}\caption{{\bf \textsf{2D maps of the FACTOR as a function of $\boldsymbol\VL$ and $\boldsymbol\VM$.~a,}}~2D voltage space showing $T^{\rm{Hahn}}_2$ as a function of gate voltages $\VM$ and $\VL$. The filled circles represent the measured coherence times $T^{\rm{Hahn}}_2$. The background is obtained by a Gaussian process interpolation and serves as a guide to the eye. {\bf \textsf{b,}}~Analogous to panel a, but for $\fR$. {\bf \textsf{c,}}~Shows the contours of $\fR$ from panel b overlaid on the plot of $\THa$ in panel a, to emphasize the overlap of both maxima.} 
\label{fig:2DMaps}
\end{figure}

\begin{spacing}{1.5}
\noindent To emphasize the overlap of maxima of $\THa$ and $\fR$, we present in Fig.~\ref{fig:2DMaps}a the 2D-voltage dependence of $\THa$, which was already introduced in the main text, as well as the 2D-voltage dependence of $\fR$ in Fig.~\ref{fig:2DMaps}b, of which only the contours were presented in the main text. Finally, Fig.~\ref{fig:2DMaps}c is obtained by overlaying the contours of the interpolated background of Fig.~\ref{fig:2DMaps}b on top of Fig.~\ref{fig:2DMaps}a.\end{spacing}
\newpage
\clearpage
\section{Impact of Control-Axis Noise on the Coherence \texorpdfstring{$T_2^{\rm{Rabi}}$}{TEXT} of the Driven Qubit}
\begin{figure}[htb!]
\centering
\includegraphics[width=1\linewidth]{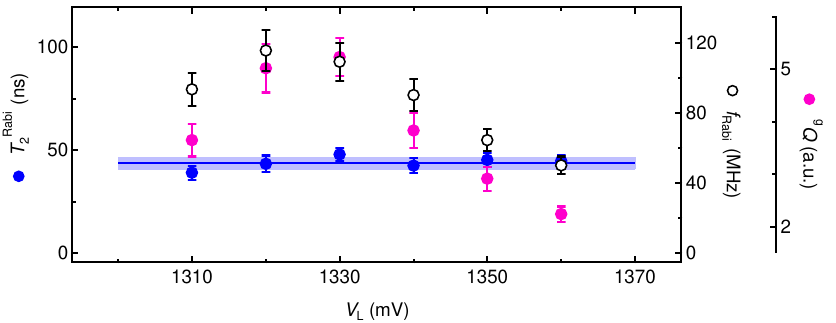} \caption{ {\bf \textsf{Coherence time $\bf \mathbf T_2^{\rm{\bf Rabi}}$ of the driven spin.}}~The blue data points show the 1/e decay time of the Rabi oscillations (left axis) as a function of the gate control parameter $V_{\rm{L}}$ over the same range as shown in Fig.~2 of the main text. The blue line highlights the average value of $\bar{T}_2^{\rm{\rm Rabi}} \approx 45$~ns and standard deviation $\sigma \approx \pm3$~ns. For reference, the Rabi frequency $f_{\rm{Rabi}}$ (white) as well as the gate quality factor $^{\rm{g}}Q = f_{\rm{Rabi}} \cdot T_2^{\rm{Rabi}}$ (pink) are also plotted for the corresponding gate voltages (right axes). Error bars for $T_2^{\rm{Rabi}}$ and $\fR$ correspond to standard deviations that result from fitting. These errors were propagated to obtain the error bars for $^{\rm{g}}Q$.} 
\label{fig:ControlAxNoise}
\end{figure}

\begin{spacing}{1.5}
\noindent When a spin qubit is continuously driven, decoherence from static dephasing is mitigated due to the reduced time spent idling in the equatorial plane. Consequently, noise on the control line becomes more significant.\\
In the rotating frame of the driven qubit a new basis can be defined known as the dressed basis \cite{Laucht2016,Laucht2016a}. In this context, the electromagnetic driving field coherently interacts with the 2-level system of the hole spin, such that the eigenstates of the driven system are no longer the $\ket{0}$ and $\ket{1}$ states, but instead the symmetric and antisymmetric superpositions of the spin states entangled with the photons of the driving field. These states constitute a new quantization axis in the driven or {\it dressed} basis where the two basis states are separated by the Rabi frequency. This change of basis of the main quantization axis shifts the noise spectral function weight towards the Larmor frequency \cite{Bosco2023}.\\
The decoherence sweet spot is associated with quasi-static dephazing in the equatorial plane and the direction of the externally applied DC-electric field for tuning the {\it{FACTOR}} is defined in relation to the main quantization axis in the laboratory frame, commonly set by the external magnetic field. Because the main quantization axis is rotated by $\pi/2$ relative to the original quantization axis, it is conceivable that the decoherence sweet spot has no significant effect in the rotating frame of the driven spin.\\
For completeness, we have plotted in Fig.~\ref{fig:ControlAxNoise} below the $T_2^{\rm{Rabi}}$ data versus the control voltage $V_{\rm{L}}$ in the range over which we show the $T_2^{\rm{Hahn}}$ data points and observe no significant dependence over that range, {\it{i.e.}} there seems to be no observable extremum in $T_2^{\rm{Rabi}}$. Provided an almost constant $T_2^{\rm{Rabi}} \approx 45$~ns, if we additionally take into account the Rabi frequencies from Fig.~1 of the Main text, the voltage dependence of the gate quality factor defined as $^{\rm{g}}Q = f_{\rm{Rabi}} \cdot T_2^{\rm{Rabi}}$, would be solely dependent on the voltage dependence of $f_{\rm{Rabi}}$ and hence the SOI. This would yield gate quality factors ranging from $\sim 2$ ($f_{\rm{Rabi}} \approx 40$~MHz) to $\sim 5$ ($f_{\rm{Rabi}} \approx 120$~MHz), and would still yield an extremum of $^{\rm{g}}Q$ peaking at the same gate voltage as $f_{\rm{Rabi}}$.\\
The experimental parameters and gate voltages $\VM$ and $\VR$ were set to the same values as for the data presented in Fig.~1a of the Main Section and are reported in the Methods Section.
\end{spacing}

\clearpage
\bibliography{BIBTEX_Carballido_etal_HotQubit_Manuscript_Temp.bib}
%

%
%